\documentclass{pasj02}
\usepackage{amsmath}
\usepackage{bm} 
\usepackage{multirow}
\usepackage{xcolor}
\usepackage{natbib}
\usepackage[singlelinecheck=false]{caption}

\newcommand{\cs}{c_{\rm s}}

\jyear{2026}
\Received{2025/07/26}
\Accepted{2026/02/03}

\begin{document}
\sloppy
\flushbottom

\title{Early Planet Formation in Embedded Disks (eDisk). XVIII. Indication of a Possible Spiral Structure in the Dust Continuum Emission of the Protostellar Disk around IRAS 16544-1604 in CB 68}
\author{
 Sanemichi Z. \textsc{Takahashi},\altaffilmark{1,2}$^{,\dag}$\orcid{0000-0003-3038-364X}\email{sanemichi.takahashi@gmail.com}
 Shigehisa \textsc{Takakuwa},\altaffilmark{1,3}\orcid{0000-0003-0845-128X}\email{takakuwa@sci.kagoshima-u.ac.jp}
 Ryosuke \textsc{Nakanishi},\altaffilmark{4}
 Yusuke \textsc{Tsukamoto},\altaffilmark{1}\orcid{0000-0001-6738-676X}\email{tsukamoto.yusuke@sci.kagoshima-u.ac.jp}
 Kazuya \textsc{Saigo},\altaffilmark{1,2}\orcid{0000-0003-1549-6435}\email{saigokz@gmail.com}
 Miyu \textsc{Kido},\altaffilmark{1,3}\orcid{0000-0002-2902-4239}\email{k1530218@kadai.jp}
 Nagayoshi \textsc{Ohashi},\altaffilmark{3}\orcid{0000-0003-0998-5064}\email{ohashi@asiaa.sinica.edu.tw}
 Zhi-Yun \textsc{Li},\altaffilmark{5}\orcid{0000-0002-7402-6487}\email{zl4h@virginia.edu}
 Leslie W. \textsc{Looney},\altaffilmark{6}\orcid{0000-0002-4540-6587}\email{lwl@illinois.edu}
 Zhe-Yu Daniel \textsc{Lin}, \altaffilmark{5}\orcid{0000-0001-7233-4171}\email{zdl3gk@virginia.edu}
 Mayank \textsc{Narang}, \altaffilmark{3}\orcid{0000-0002-0554-1151}\email{mayankn1154@gmail.com}
 Kengo \textsc{Tomida}, \altaffilmark{7}\orcid{0000-0001-8105-8113}\email{tomida@astr.tohoku.ac.jp}
 John J. \textsc{Tobin}, \altaffilmark{8}\orcid{0000-0002-6195-0152}\email{jtobin@nrao.edu} 
 Jes K. J{\o}rgensen \altaffilmark{9}\orcid{0000-0001-9133-8047}\email{jesk.jorgensen@gmail.com}
}

\maketitle

\altaffiltext{1}{Department of Physics and Astronomy, Graduate School of Science and Engineering, Kagoshima University, 1-21-35 Korimoto, Kagoshima, Kagoshima 890-0065,
Japan 181-8588, Japan}
\altaffiltext{2}{National Astronomical Observatory of Japan, 2-21-1 Osawa, Mitaka, Tokyo 181-8588, Japan}
\altaffiltext{3}{Academia Sinica Institute of Astronomy \& Astrophysics,
11F of Astronomy-Mathematics Building, AS/NTU, No.1, Sec. 4, Roosevelt Rd,
Taipei 10617, Taiwan}
\altaffiltext{4}{Physics and Astronomy Program, School of Science, Kagoshima University, 1-21-35 Korimoto, Kagoshima, Kagoshima 890-0065,
Japan 181-8588, Japan}
\altaffiltext{5}{University of Virginia, 530 McCormick Rd., Charlottesville, Virginia 22904, USA}
\altaffiltext{6}{Department of Astronomy, University of Illinois, 1002 West Green St, Urbana, IL 61801, USA}
\altaffiltext{7}{Astronomical Institute, Graduate School of Science, Tohoku University, Sendai 980-8578, Japan}
\altaffiltext{8}{National Radio Astronomy Observatory, 
520 Edgemont Rd., Charlottesville, VA 22903 USA}
\altaffiltext{9}{Niels Bohr Institute, University of Copenhagen,
{\O}ster Voldgade 5-7, 1350, Copenhagen K, Denmark}


\clearpage

\begin{abstract}
We performed numerical simulations along with radiative transfer calculations
to reproduce an intriguing asymmetric shoulder feature in the dust continuum emission
of the protostellar disk around one of the eDisk targets,
Class 0 protostar IRAS 16544-1604 in CB 68. This is our first attempt
to bridge between theoretical works of protostellar disk evolution and the eDisk observations.
We found that while our hydrodynamic simulations form
spiral structures caused by gravitational instability,
they become less discernible after the disk is inclined and convolved with the telescope beam.
The widths of the spiral structure as obtained by our numerical simulations
are $\sim$0.1 -- 0.8 times the eDisk beam size of 4.5 au.
Our modeling effort implies that the apparent absence of spiral features in the eDisk observations does not necessarily indicate the real absence of internal substructures and gravitational instability.
We also found that the asymmetric shoulder structure of the continuum profile along the major axis appears when the disk is massive enough with a Toomre parameter $Q\sim 1$.
This mechanism offers a potential explanation for the observed, asymmetric shoulder features in the disks surrounding IRAS 16544-1604
and the other eDisk sources.
\end{abstract}

\section{Introduction} \label{sec:intro}
\footnotetext[$\dag$]{Present address: Matsuda Company}

\ \ \ \ Protoplanetary disks are the birthplaces of planets. 
Since the physical conditions of protoplanetary disks profoundly influence the planet formation processes \citep{1997Sci...276.1836B, 2005A&A...430.1133K, 2012A&A...541A..97M,2015ApJ...814..130M, 2015ApJ...798..112M, 2018AJ....156..221N},
a comprehensive understanding of the disk properties is crucial for developing planet formation theories.
High resolution observations of protoplanetary disks in Class II sources reveal that
the dust continuum emission in the disks is not smooth but shows intriguing structures, such as spiral arms \citep[$e.g.,$][]{2016Sci...353.1519P}, crescents \citep[$e.g.,$][]{2013Sci...340.1199V, 2013PASJ...65L..14F}, and ring-gap structures \citep[$e.g.,$][]{2015ApJ...808L...3A, 2016ApJ...820L..40A, 2016PhRvL.117y1101I, 2017ApJ...840L..12S, 2017A&A...600A..72F, 2018A&A...610A..24F}.
Especially, Disk Substructures at High Angular Resolution Project (DSHARP) \citep{2018ApJ...869L..41A} observations have found that ring-gap structures are common
in protoplanetary disks, which are considered to be carved by the hidden protoplanets.
On the contrary, our ALMA Large Program, 
"Early Planet Formation in Embedded Disks (eDisk)", has found that 
majority of protostellar disks around Class 0 / I sources do not exhibit such ring-gap features, except for the disks around the most evolved Class I sources of L1489 IRS \citep[][]{2023ApJ...954...67S} and Oph IRS 63 \citep[][]{2023ApJ...958...98F}.
Instead, in many of the eDisk sources (L1527 IRS, IRAS 16253-2429, GSS 30 IRS3,  IRAS 16544-1604, R CrA IRAS 32B, IRAS 04302+2247, and R CrA IRS7B-a), the 1.3-mm dust images exhibit asymmetric intensity profiles along the disk minor axes. Our radiative transfer modeling has revealed that the asymmetric intensity profiles along the disk minor axes can be interpreted as the effect of the dust flaring
and optically-thick dust emission \citep{2023ApJ...951....8O, 2024ApJ...964...24T}.

Several eDisk sources also exhibit intriguing shoulder, bump,
or asymmetric features along the disk major axes.
In L1527 IRS, a Class 0 protostar in Taurus, the southern part of the north-south elongated disk feature is stronger and wider
along the transverse direction than the northern part
\citep{2022ApJ...934...95S,2023ApJ...951...10V}. Toward Ced110 IRS4A, a Class 0 source in the Chamaeleon I region, bump features are clearly seen at radii of 30-70 au, with a slight asymmetry
along the major axis.
Three intensity peaks to the west, center, and east are seen along the disk major axis toward Ced110 IRS4B, where the western peak is slightly stronger than the eastern peak \citep{2023ApJ...954...67S}. A Class 0 source of IRAS 16253-2429 in Ophiuchus clearly exhibits an asymmetric intensity profile along the disk major axis \citep{2023ApJ...954..101A}. Toward the Class 0 protostar R CrA IRS5N \citep{2023ApJ...954...69S} and Class I protostar R CrA IRS 7B-a \citep{2024ApJ...964...24T}, the locations of the emission peaks are shifted from the geometrical disk center along the major axes. Double-side, shoulder features are identified in the protostellar disk of GSS 30 IRS 3 at radii of ~25 and 50 au \citep{2024A&A...690A..46S}, while an one-side shoulder is seen toward IRAS 16544-1604 at a radius of $\sim$15 au \citep{2023ApJ...953..190K}, both of which are Class 0 sources in Ophiuchus. These results imply that protostellar disks have internal, possibly non-axisymmetric, substructures. The nature of such internal structures is, however, yet to be elucidated.

According to numerical simulations of star and disk formation, young protostellar disks can be massive enough to be gravitationally unstable \citep[$e.g.,$][]{1998ApJ...508L..95B,2010ApJ...718L..58I}.
In such a gravitationally unstable disk, spiral arms should be formed,
which play an important role in angular momentum transfer
\citep[$e.g.,$][]{2011PASJ...63..555M,2015MNRAS.452..278T}.
Absence of such spiral features implies that these disks are gravitationally stable,
unless such features appear only in a short time scale for observations.
The source of the bump or shoulder features observed by eDisk
could be such spiral structures
hidden in the protostellar disks.
To investigate the origin and physics of the observed bump / shoulder features,
hydrodynamic simulations coupled with radiative transfer calculations
are essential.

\begin{figure}
    \centering
    \includegraphics[width=9.5cm]{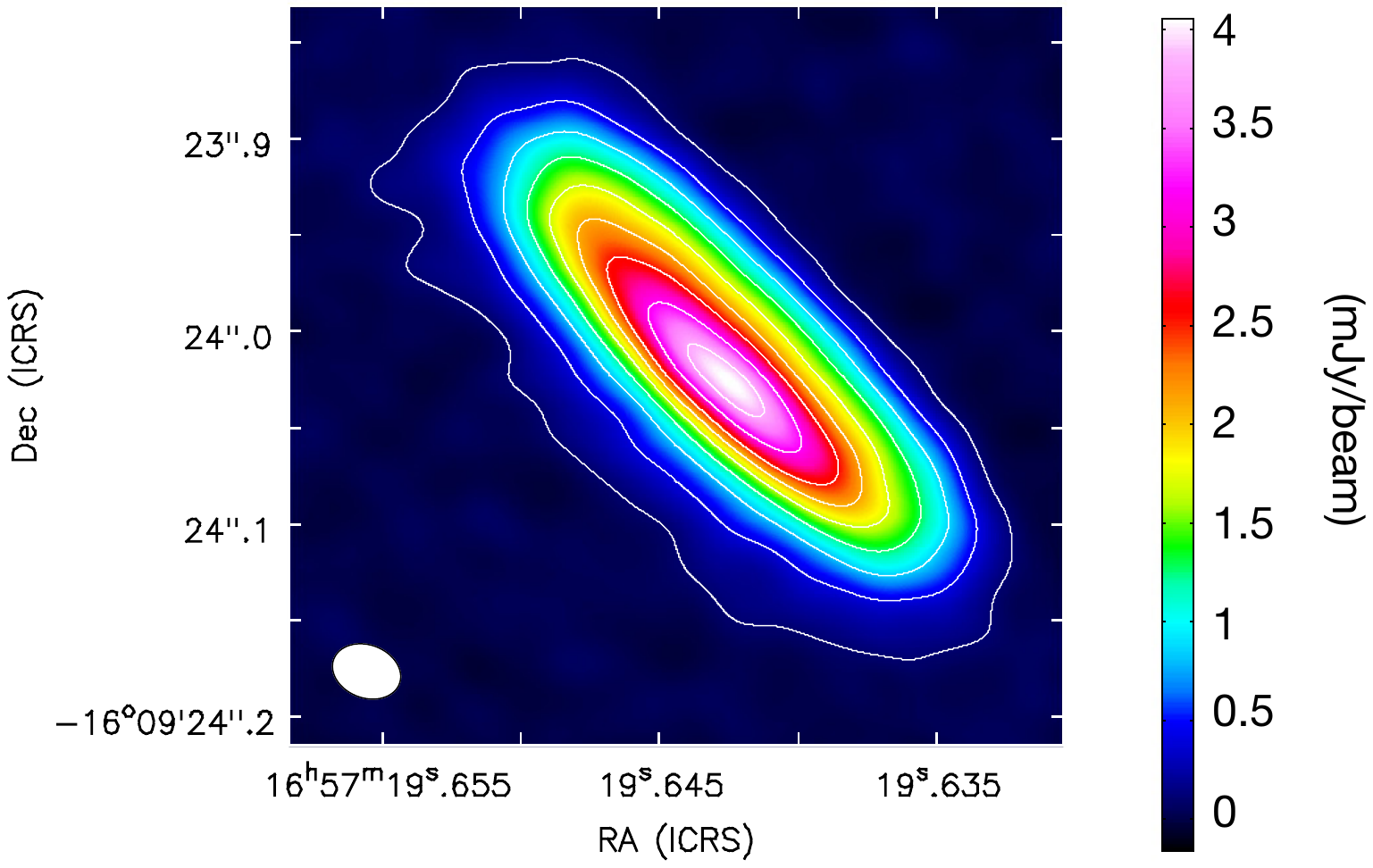}
    \caption{1.3 mm dust continuum image in IRAS 16544 obtained by eDisk \cite[]{2023ApJ...953..190K}. 
    Contour levels are 5$\sigma$, 20$\sigma$, 40$\sigma$, 60$\sigma$, 80$\sigma$, 100$\sigma$, 120$\sigma$, and 180$\sigma$ (1$\sigma$ = 21 $\mu$Jy beam$^{-1}$). The beam size is 0\farcs036$\times$0\farcs027 (P.A = 69 degree), as shown in a filled ellipse at the bottom-left corner. {Alt text: Color and contour image.}}
    \label{fig:obs_image}
\end{figure}

\begin{figure*}
    \centering
    \includegraphics[width=16cm]{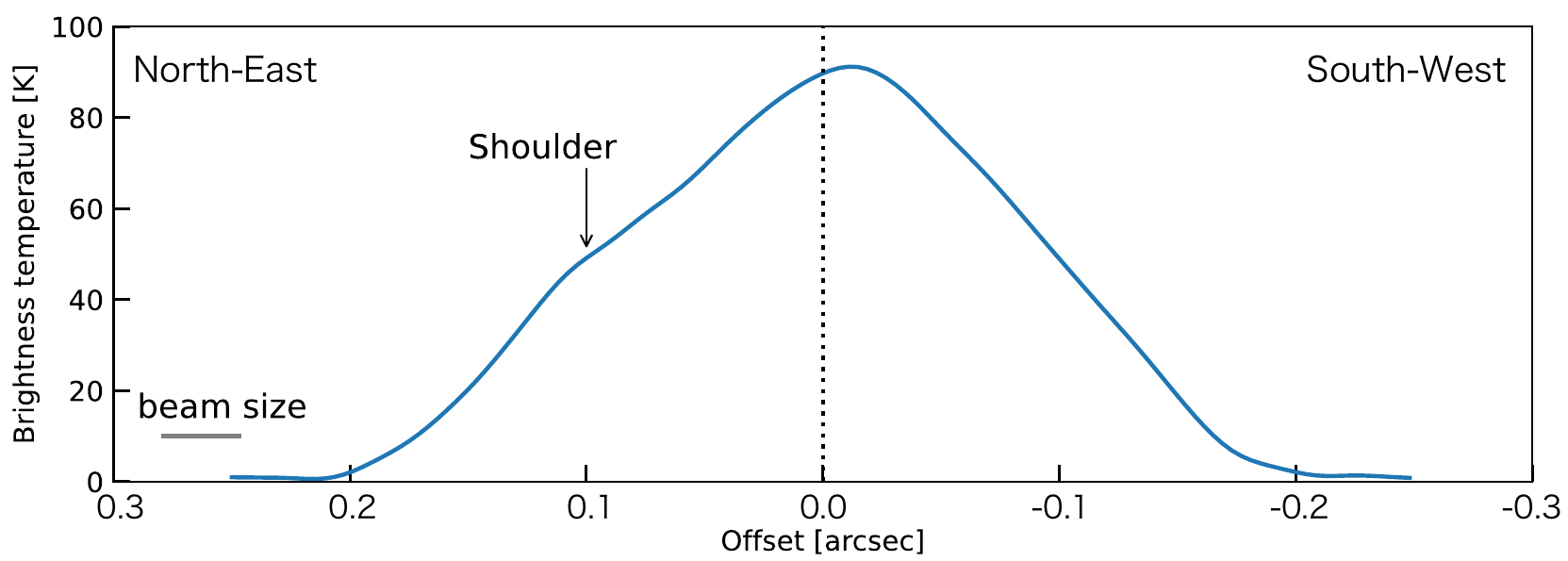}
    \caption{Observed brightness temperature profile of the 1.3 mm dust continuum emission along the disk major axis in IRAS 16544. The beam size is shown in the bottom left.
    A shoulder structure is observed in the left-hand side of this profile.
    {Alt text: Line graph. x axis shows the positional offset from the protostellar
    position from 0.3 to -0.3 arcsecond. The y axis shows the brightness temperature
    in Kelvin.}
    }
    \label{fig:obs_major}
\end{figure*}

This paper is the first such attempt to compare results of hydrodynamical simulations
plus radiative transfer calculations to the eDisk images. 
Such a comparison requires a number of model parameters and their fine tuning,
and it is impractical to conduct such a task for all the eDisk
images at once.
We focus on disk structures around IRAS 16544-1604 (hereafter IRAS 16544),
which exhibits a clear shoulder feature along the disk major axis.
This object is located in the southwestern outskirt of the Ophiuchus
North region at a distance of 151 pc  \citep{2020A&A...633A..51Z},
and its bolometric luminosity and central protostellar mass are estimated to be
0.89 $L_\odot$ \citep[]{2023ApJ...951....8O} and $0.14\ M_{\odot}$
\citep[]{2023ApJ...953..190K}
\footnote{This dynamical mass is obtained from the fitting of
the C$^{18}$O (2--1) ridge emission in the Position - Velocity diagram.},
respectively.
Figure \ref{fig:obs_image} shows the 1.3 mm dust continuum image of the protostellar disk
in IRAS 16544 taken from \cite{2023ApJ...953..190K}.
The observed dust disk radius, which is determined by the region where the dust continuum intensity
exceeds 5 $\sigma$ (1$\sigma$ = 21 $\mu$Jy beam$^{-1}$), is $\sim$30 au.
While there is no clear spiral structure seen in the image,
the brightness temperature profile along the disk major axis
shows an asymmetry with a ``shoulder'' structure (Figure \ref{fig:obs_major})
\footnote{The origin and the major axis for the profile are
determined by the 2-dimensional Gaussian fitting to the
observed 1.3-mm image. Through the radiative transfer calculations,
we have verified that the simple 2-dimensional Gaussian fitting is sufficient
to derive the central protostellar position and the major axis,
even if the 3-dimensional disk structure has a flared geometry
(see Appendix A).}.
\citet[]{2023ApJ...953..190K} estimated the total gas + dust disk mass
in the range from
1.63 $\times$ 10$^{-3}$ $M_{\rm \odot}$ to
1.02 $\times$ 10$^{-2}$ $M_{\rm \odot}$
for the assumed dust temperature of 20 - 100 K.
These estimates should be regarded as the lower limit of the disk mass because
of the assumption of the optically thin dust emission.
This gives the upper limit of the Toomre Q value of
2.7 - 17.
Therefore, it is possible that the disk is gravitationally unstable.

We calculated density and temperature distributions of gravitationally unstable disks by numerical simulations and evaluated the intensity distributions of the 1.3 mm continuum emission arising from the disks using radiative transfer calculations.
Through the comparison between the intensity profiles obtained from theoretical calculations and the observations, we investigated the disk substructure that can explain the observations,
specifically, the asymmetric shoulder structure observed by eDisk.
Although the comparison between the spiral structures observed around Elias 2–27 \cite[]{2016Sci...353.1519P} and produced by numerical simulations of the gravitationally unstable disk has already been presented in \cite{2017ApJ...835L..11T}, the detectability of the spiral arms is not well studied so far.
We also discuss the requirement to observationally detect the spiral arms in the protostellar disk based on the physical values obtained from the numerical simulations.

This paper is organized as follows. In Section 2, the setup of the numerical simulations of the self-gravitating disk and the radiative transfer calculations is presented.
Section 3 shows the results of the numerical simulations and the radiative transfer calculations.
Sections 4 and 5 provide a few key insights obtained from our work and conclusions, respectively.

\section{Method}\label{sec:method}

\subsection{Basic Equations of Numerical Simulation}\label{sec:fargo}

\ \ \ \ A grid-based two-dimensional hydrodynamical simulation code, \textsc{fargo-adsg}, is adopted to calculate the disk gas dynamics and spiral structure formation through the gravitational instability.
The equations solved in the simulation are as follows:
\begin{equation}
    \frac{\partial \Sigma}{\partial t}+\nabla\cdot(\Sigma \bm{v})=0
\end{equation}
\begin{equation}
\Sigma\left( \frac{\partial \bm{v}}{\partial t} + \bm{v}\cdot\nabla \bm{v} \right)
= -\nabla P -\nabla \Phi
    \label{eq:eom}
\end{equation}
\begin{equation}
    \frac{\partial E}{\partial t} + \nabla(E \bm{v}) = -P\nabla\cdot \bm{v} -\Lambda_{\rm c},
    \label{eq:eoe}
\end{equation}
where $\Sigma$ is the gas surface density, $\bm{v}$ is the gas velocity, $E$ is the gas internal energy per unit area, $P$ is the vertically integrated pressure, $\Phi$ is the gravitational
potential, and $\Lambda_{\rm c}$ is the net cooling rate per unit area. 
Although the heating term is not included in this equation explicitly, the effects of the radiation and accretion heating are included in the net cooling rate (see the explanation of Equation \ref{eq:Lambda_c} below).
$P$ is calculated assuming an ideal gas equation of state with the thin disk approximation,
\begin{equation}
    P=(\gamma-1)E,
\end{equation}
where $\gamma = 5/3$ is the ratio of specific heat.
Gas and dust are assumed to be at the same temperature and the temperature $T$ is calculated from 
\begin{equation}
    T=\frac{\mu m_{\rm H}}{k_{\rm B}}\frac{P}{\Sigma},
\end{equation}
where $\mu~(=2.34)$, $m_{\rm H}$, and $k_{\rm B}$ are the mean molecular weight, the hydrogen mass, and the Boltzmann constant, respectively.

The viscous term is not included in the equation of motion (Equation \ref{eq:eom}), as viscosity does not play an important role in spiral structure formation. Spiral arms due to gravitational instability form and disperse recurrently within the dynamical timescale. In contrast, the viscous timescale is much longer than the dynamical timescale. When the width of the spiral is similar to the pressure scale height (see Section \ref{beamsize}), the viscous timescale can be estimated as $\sim 1/(\alpha \Omega)$, where $\alpha$ is the parameter representing the strength of viscosity and $\Omega$ is the Keplerian frequency. 
For example, at a radius of 30 au, the dynamical time is $\sim 1/\Omega=70$ yr and the viscous time is $\sim 1/(\alpha \Omega)=7000$ yr on the assumption of $\alpha =0.01$.
In cases where $\alpha$ is much smaller than unity, viscosity can be considered negligible.


The net cooling rate $\Lambda_{\rm c}$ is given as follows \cite[]{1990ApJ...351..632H},
\begin{equation}
    \Lambda_{\rm c}= \frac{8}{3}\sigma(T^4-T_{\rm eq}^4)\frac{\tau}{\frac{1}{4}\tau^2+\frac{1}{\sqrt{3}}\tau +\frac{2}{3}},
    \label{eq:Lambda_c}
\end{equation}
where $\sigma$ is the Stefan-Boltzmann constant, $T_{\rm eq}$ is the equilibrium temperature, and $\tau$ is the optical depth of the dust continuum emission.
As will be shown below the 1.3-mm dust-continuum emission is optically thick.
The distribution of the equilibrium temperature ($\equiv T_{\rm eq}$) in the disk is set to roughly reproduce
the observed 1.3-mm brightness temperature distribution 
without the shoulder structure as,
\begin{equation}    
    T_{\rm eq} = {\rm max}\left[ T_{\rm peak} - Ar, 10{\rm [K]}\right],
    \label{eq:Teq}
\end{equation}
where $r$ is the radius from the central star, $T_{\rm peak}=100$ K is the peak temperature of the disk consistent with the observed peak brightness temperature of the 1.3-mm dust-continuum emission, and $A$ = 3 $({\rm K~au^{-1}})$ is the temperature gradient of the disk estimated from the observations\footnote{{If a power law distribution
of $T_{\rm eq}$ is adopted as commonly used in the disk temperature distribution, the temperature around the central region becomes significantly higher than the observed brightness temperature. Therefore, for comparison with observations, we use the observed brightness temperature to model $T_{\rm eq}$. If the temperature structure in the z-direction is calculated directly, the temperature distribution at the $\tau\sim1$ surface may closely match the observed brightness temperature, even with a power law temperature distribution in the midplane. However, since this 2-dimensional study cannot handle the temperature structure in the z-direction, we adopt
this approximate method.}} \cite[]{2023ApJ...953..190K}.
The optical depth $\tau=\kappa_{\rm R} \Sigma$ is
obtained using the Rosseland mean opacity per unit gas mass (= $\kappa_{\rm R}$) calculated from the opacity model given by \cite{2018ApJ...869L..45B} with the minimum and maximum dust sizes\footnote{Since the maximum dust size in protostellar disks is still controversial, it is assumed to be 1 mm in this work.
Numerical simulations have reproduced dust growth up to 1~mm
in protostellar disks \cite[]{2023PASJ...75..835T}.} of 0.1 ${\rm \mu m}$ and 1 mm and a power-law index of the grain size distribution of -3.5. The dust is assumed to be dynamically
and thermally coupled with gas, and the gas-to-dust mass ratio to be 100.
If compressional heating is ineffective in the disk (when the first term on the right-hand side of Equation \ref{eq:eoe} is negligible), the disk temperature evolves so that the cooling rate $\Lambda_{\rm c}$ is zero, $i.e., T = T_{\rm eq}$.
$\Lambda_{\rm c}$ represents the radiative cooling in the vertical direction of the disk when $T_{\rm eq} = 0$.
Non-zero $T_{\rm eq}$ corresponds to the addition of a heating term.
Since the temperature distribution of the disk is affected not only by the radiative cooling term but also by advection and compressional heating, the temperature distribution of the simulated disk deviates from $T_{\rm eq}$.


\subsection{Numerical Procedure and Initial Conditions}\label{sec:setup}

\ \ \ \ In the numerical simulations, the 2-dimensional polar grid ($r,\ \phi$) is adopted.
The inner and outer boundaries of the numerical simulations
are set to be 1 au and 300 au, respectively, and the open boundary condition (the gradients of the physical values are assumed to be 0 at the boundaries) is used.
The numerical simulations do not include the envelope component.
The grid numbers for the radial and azimuthal directions are 512 and 1024, respectively.
The central protostellar mass is set to be $M_*=0.14\ M_{\rm \odot}$, as derived by \cite{2023ApJ...953..190K}.
The initial temperature distribution is given by $T_{\rm ini}=T_{\rm eq}$.
The initial surface density is given as
\begin{equation}
    \Sigma_{\rm ini} = \Sigma_{\rm 1au}\left(\frac{r}{1~{\rm au}}\right)^{-1.3}
    \sqrt{\frac{T_{\rm ini}}{97~{\rm K}}}\exp\left(-\frac{r}{r_{\rm disk}}
\right),
\label{eq:sigmaini}
\end{equation}
where $\Sigma_{\rm 1au}$ is the surface density at $r=1$ au,
and $r_{\rm disk}~(\gg 1~{\rm au})$ is the outer disk radius determined by
the exponential taper.
The radial exponent $-1.3$ and the $\sqrt{T_{\rm ini}}$ term are adopted to keep the Toomre's Q values \cite[]{1964ApJ...139.1217T} approximately constant in the inner region of the disk. This initial surface density profile is expected from hydrodynamic simulations of disk formation \cite[]{2015MNRAS.446.1175T}. The term $\sqrt{T_{\rm ini}/97~{\rm K}}$ cancels the dependence of $Q$ values on the sound speed, where 97 K is the initial temperature at $r=1$ au.

\begin{figure}
    \centering
    \includegraphics[width=9cm]{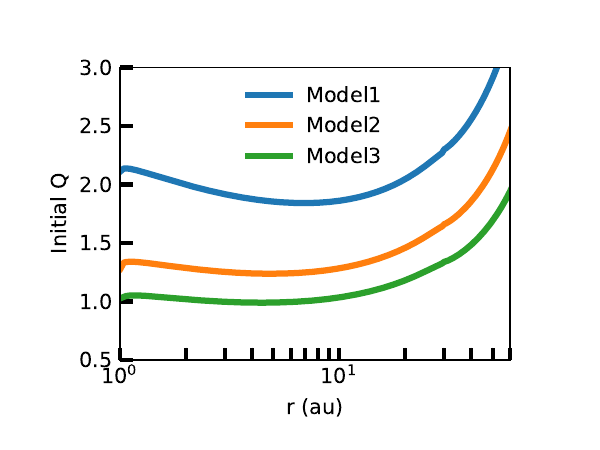}
    \caption{Radial distribution of initial $Q$ values for Models 1, 2 and 3.
    {Alt text: Line graph. x axis shows the radius from 1 to 60 astronomical unit
    in log scale. The y axis shows the Toomre $Q$ value from 0.5 to 3 in linear scale.}}
    \label{fig:initialQ}
\end{figure}

A small level (=0.1 per cent) of white noise is added to the initial surface density as an initial perturbation of the gravitational instability.
The initial radial velocity is zero, and the initial azimuthal velocity is provided to balance the centrifugal and pressure gradient force with the gravitational force.
Three models (Models 1, 2, and 3)
with different initial disk surface densities and thus disk masses are adopted.
The adopted values of $\Sigma_{\rm 1au}$ and disk masses
are listed in Table \ref{tab:param}.
The disk mass of Model 1 is smaller than the central protostellar mass,
that of Model 2 is comparable to the protostellar mass,
and that of Model 3 is higher than the protostellar mass.
These three models yield 
the initial value of $Q=c_{\rm s}\Omega/(\pi G \Sigma)$ of $\sim$2,
1.3, and 1.0, respectively (Figure \ref{fig:initialQ}),
where $c_{\rm s}$ is the sound speed and $\Omega$ is the angular velocity.
The disk outer radius is set as $r_{\rm disk}=60 $ au, which is the outermost radius where the Keplerian rotation is observed in the
C$^{18}$O (2--1) emission \citep{2023ApJ...953..190K}.
Parameters used for the disk models are summarized in Table \ref{tab:param}.

With these initial conditions, evolutions of
the surface density and (vertically-averaged) temperatures are
calculated through numerical simulations.
For Models 1 and 2, numerical simulations are continued
until $t = 2.4\times 10^{4}$ yr.
This duration is long enough for the growth of spiral arms due to
the gravitational instability.
For Model 3, the speed of the numerical simulation becomes
much slower than the other two models, due to the higher
disk surface density and gravitational instability.
The results of Model 3 at $t = 3.5\times 10^{3}$ yr
are presented, which corresponds to the 8 orbital period
at $r =$30 au. At this time three spiral, $m = 3$ feature
has been developed. On the other hand, after
$t \sim 3.9\times 10^{3}$ yr the model disk is fragmented
to show a binary feature, which is not applicable to the present
case of IRAS 16544.
The surface density and temperature
distributions are then input to the radiative transfer calulations
with RADMC3d described below.

\begin{table*}[t]
\caption{Summary of the model parameters}
\label{tab:param}
    \centering
\begin{tabular}{lll}
\hline
\hline
\multicolumn{3}{c}{Model parameters adopted from \cite{2023ApJ...953..190K} }\\
\hline
Central protostellar mass & $M_*$ & 0.14 $M_{\rm \odot}$\\
Peak temperature & $T_{\rm peak} $& 100 [K] \\
Temperature gradient & $A$ & 3 [K au$^{-1}$] \\
Disk outer radius & $r_{\rm out }$ & 60 [au]\\
Dust disk radius & $r_{\rm dust}$ & 30 [au] \\
Beam size for Gaussian convolution && 4.5 au\\
\hline
\multicolumn{3}{c}{The other parameters}\\
\hline
\multirow{2}{*}{Surface density at 1 au} & \multirow{2}{*}{$\Sigma_{\rm 1au}$ }
&$1.2\times 10^4\ [{\rm g~cm^{-2}}]$ (Model 1)  \\
& & $2.1\times10^4 \ [{\rm g~cm^{-2}}]$ (Model 2)\\
& & $2.5\times10^4 \ [{\rm g~cm^{-2}}]$ (Model 3)\\
\multirow{2}{*}{Initial disk gas mass} & & {0.1 $M_{\odot}$} (Model 1)\\
&&{0.17} $M_{\rm \odot}$ (Model 2)\\
&&{0.21 $M_{\rm \odot}$ (Model 3)}\\
Radial exponent of the surface density profile &&$-1.3$\\
Dust-to-gas mass ratio & $\epsilon$ & 0.01\\
Opacity at 1.3 mm & $\kappa_{\rm 1.3mm}$ & 1.88 ${\rm cm^2~g^{-1}}$\\
\hline
\end{tabular}
\end{table*}

\subsection{Radiation Transfer Calculations}
\label{sec:dustmodel}

\ \ \ \ To perform radiative transfer calculations, 
density and temperature distributions in the 3-dimensional space
need to be constructed from the 2-dimensional hydrodynamic results
with FARGO-ADSG. On the assumption of the vertically isothermal
condition, 
the 2-dimensional temperature distribution $T(r,\ \phi)$
is converted into the three dimensional temperature
distribution $T(r,\ \phi,\ z)$. Whereas the vertical isothermal
condition is over-simplified, it is sufficient for
the purpose of the present study that
compares the observed and model dust-continuum images
and investigates the possible source of the observed shoulder
feature.
The vertical gas density distribution is assumed to be Gaussian, and the formulation of the scale height $H$ by \cite{1999A&A...350..694B} is adopted, which is suitable for self-gravitating massive disks,
\begin{equation}
H (r,\phi)=\frac{\cs^2}{\pi G \Sigma}\left(\frac{\pi}{4Q^2}\right)
\left[\sqrt{1+\frac{8Q^2}{\pi}}-1\right].
\label{eq:Hmodel}
\end{equation}

The observed extent of the dust continuum emission is
$r_{\rm dust}~\sim$30~au above the 5$\sigma$ level,
while the gas disk radius is $r_{\rm out}~\sim$60~au.
We found that the same radius for both the dust and gas disks
makes the spatial extent of the model continuum image too large
compared to that of the observed image.
An exponential cutoff of the dust density is thus added as
\begin{equation}
    \rho_{\rm dust}(r,\phi,z) = \epsilon \frac{\Sigma_{\rm gas} (r,\phi)}{\sqrt{2\pi}H(r,\phi)}\exp{\left(-\frac{r}{r_{\rm dust}}\right)}\exp{\left(-\frac{1}{2}\left(\frac{z}{H(r,\phi)}\right)^2\right)},
\end{equation}
where $\Sigma_{\rm gas} (r,\phi)$ denotes the gas surface density
obtained from the \textsc{fargo-adsg} simulations, and
$\epsilon=0.01$ is the dust-to-gas mass ratio.
While this is an additional, ad-hoc procedure, this
manipuation does not affect results and physics of the numerical
simulations of the spiral-arm formation.
It is not straightforward to reproduce
the different gas and dust radii with numerical simulations.
A smaller dust radius than the gas radius has also been
found in another eDisk target, R CrA IRS7B-a \cite[]{2024ApJ...964...24T}.

With the 3-dimensional density and temperature distributions
given above, radiative transfer calculations using RADMC-3D \cite[]{2012ascl.soft02015D} are performed.
The disk inclination angle is set to be 73$^\circ$ as obtained from the observations.
The wavelength of the radiation transfer is 1.3 mm for the comparison with the eDisk observations.
The dust model used to calculate the opacity at 1.3 mm is based on the model of the DSHARP collaboration \cite[]{2018ApJ...869L..45B}, which assumes a dust size distribution with the minimum and maximum dust sizes of 0.1 $\mu$m and 1 mm, respectively, and a power-law index of -3.5. This is the same model used to calculate the Rosseland mean opacity in the hydrodynamic simulation. The resultant opacity at 1.3 mm is $\kappa_{1.3\ \rm mm} = 1.88\ \rm [cm^2\ g^{-1}]$.
After the radiative transfer calculations, the resultant model images are convolved by a Gaussian with a FWHM of 4.5 au, which approximately corresponds to the beam size of the eDisk observations.

\section{Results}\label{sec:results}
\subsection{Model 1: Low Surface Density Case}
\label{ft}

\begin{figure*}
    \includegraphics[width=8.8cm]{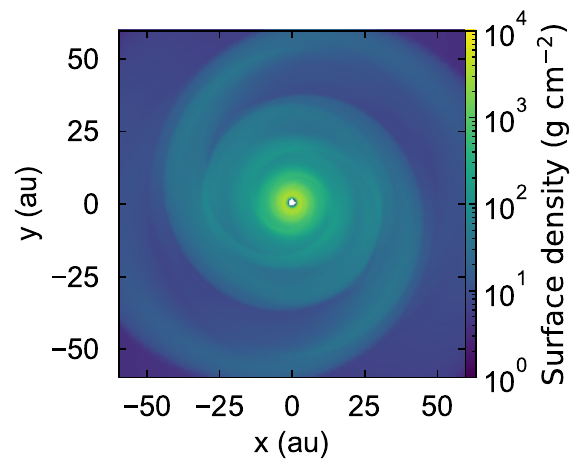}
    \includegraphics[width=8.8cm]{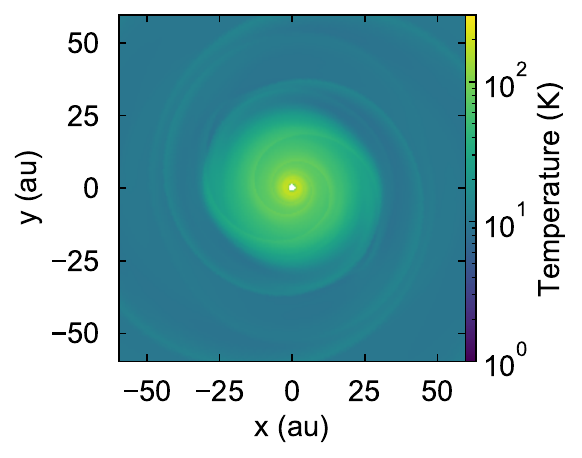}    
    \caption{Results of the numerical simulation of the self-gravitating disk of Model 1 at
    $t \sim$ $2.4\times 10^{4}$ yr after the start of the simulation. The left and right panels show the surface density and the temperature images, respectively. The white regions at the center of these panels correspond to the inner boundary of the simulation. The spiral structures are formed in both surface density and temperature distributions due to the gravitational instability and the compressional heating.
    {Alt text: Two color images.}
}
    \label{fig:sim_results}
\end{figure*}

\ \ \ \ The surface density and temperature distributions of Model 1,
the case of the low initial surface density, after $t =$ $2.4\times 10^{4}$ yr
are shown in Figure \ref{fig:sim_results}.
Because of the self-gravitational instability, spiral arms are formed in the density distribution.
In the spiral arms, the temperature is also increased because of the compressional heating. The spiral arms are thus seen in the temperature map.

\begin{figure*}
\begin{tabular}{cc}
\begin{minipage}[t]{0.36\hsize}
    \centering
    \includegraphics[keepaspectratio, scale=0.6]{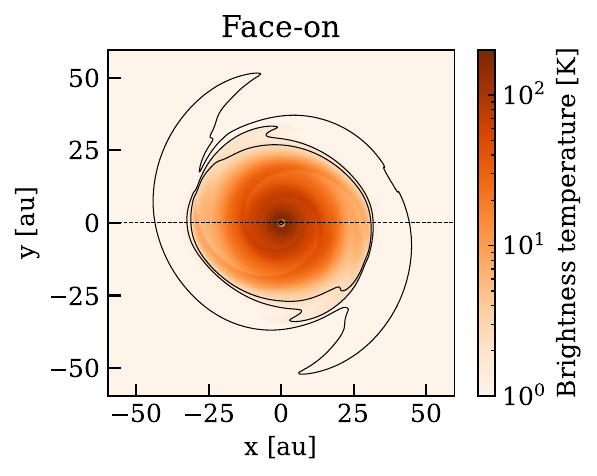}   
\end{minipage} &
\begin{minipage}[t]{0.54\hsize}
    \centering
    \includegraphics[keepaspectratio, scale=0.6]{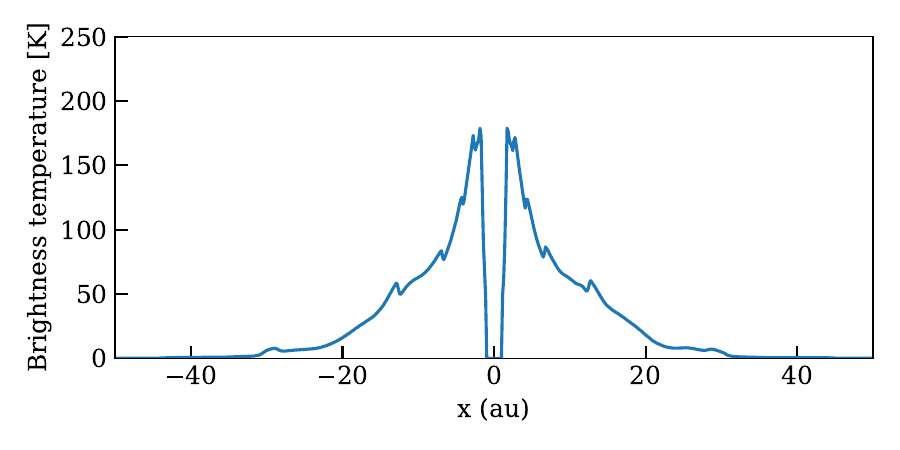}    
\end{minipage} \\
\begin{minipage}[t]{0.36\hsize}
    \centering
    \includegraphics[keepaspectratio, scale=0.6]{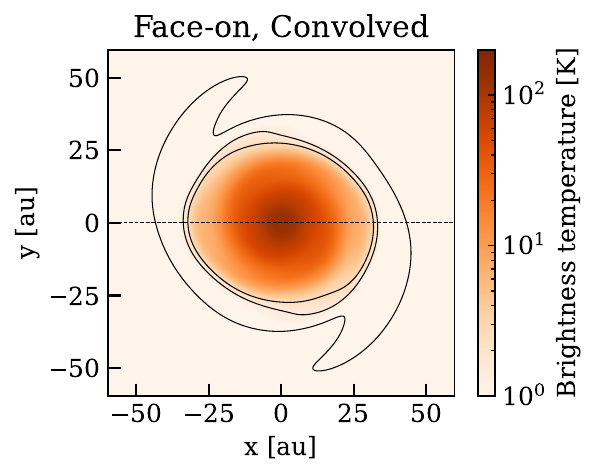}    
\end{minipage} &
\begin{minipage}[t]{0.54\hsize}
    \centering
    \includegraphics[keepaspectratio, scale=0.6]{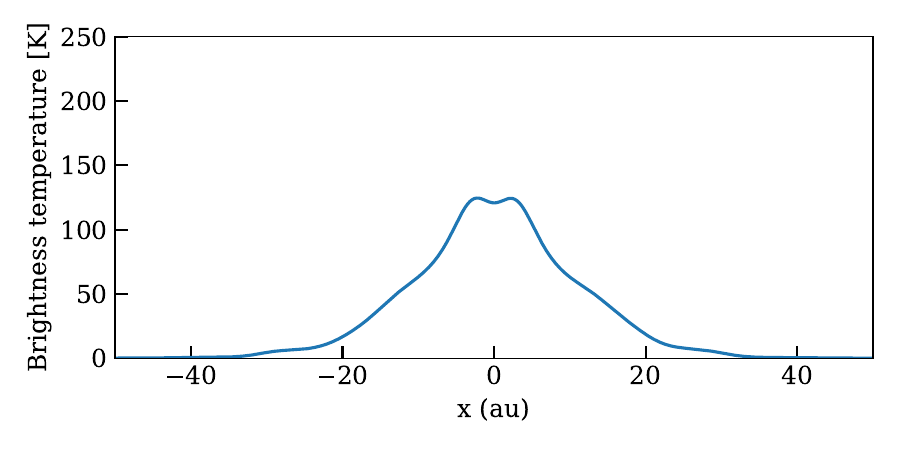}    
\end{minipage} \\

\begin{minipage}[t]{0.36\hsize}
    \centering
    \includegraphics[keepaspectratio, scale=0.6]{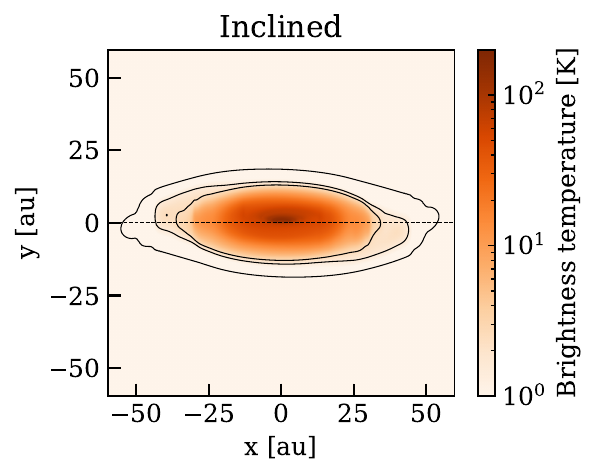}    
\end{minipage} &
\begin{minipage}[t]{0.54\hsize}
    \centering
    \includegraphics[keepaspectratio, scale=0.6]{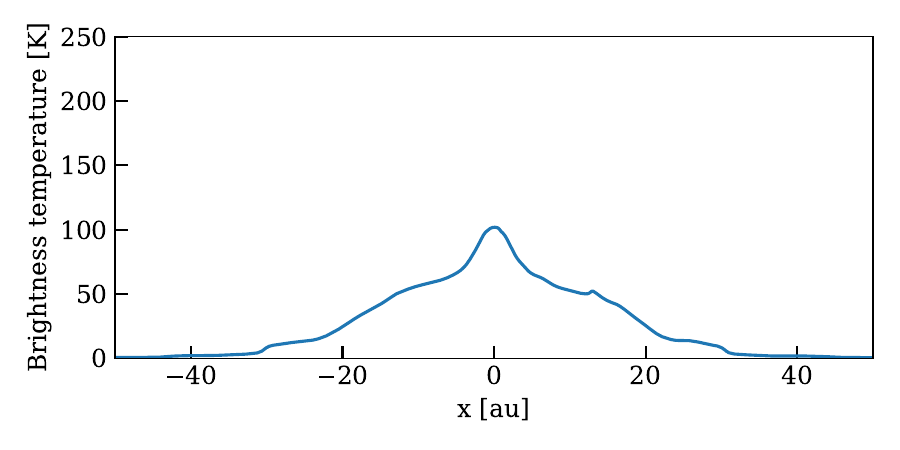}    
\end{minipage} \\

\begin{minipage}[t]{0.36\hsize}
    \centering
    \includegraphics[keepaspectratio, scale=0.6]{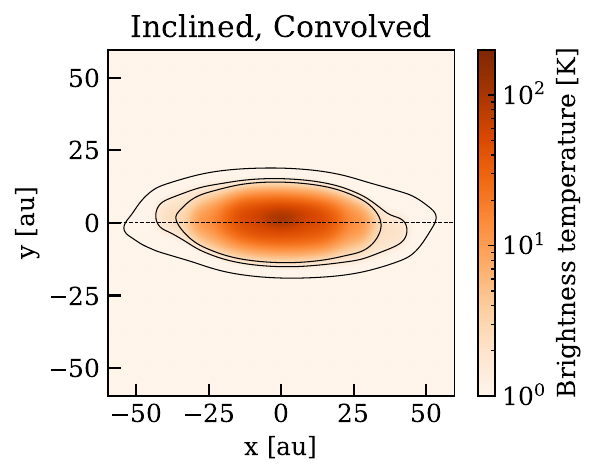}    
\end{minipage} &
\begin{minipage}[t]{0.54\hsize}
    \centering
    \includegraphics[keepaspectratio, scale=0.6]{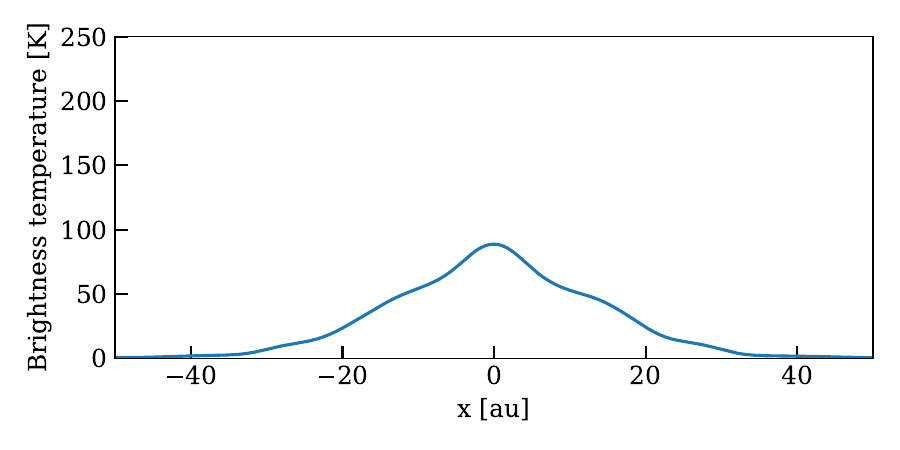}    
\end{minipage} 
\end{tabular}
\caption{
1.3-mm dust continuum images (left) and intensity profiles along the major axis of the model disk (right) obtained with the radiative
transfer calculations. Contour levels in left panels are 1$\sigma$, 3$\sigma$, and 5$\sigma$ of the eDisk observations (1$\sigma$= 0.5 K).
The corresponding
density and temperature distributions
are shown in Figure \ref{fig:sim_results}.
Dashed lines denote the cuts of the intensity profile of the relevant right panels.
The results of the face-on view without
beam convolution, the face-on view with convolution, the inclined view without convolution, and the inclined view with convolution, are presented in each row from top to bottom.
In the panels of the face-on view without convolution (the first row), the brightness temperature is decreased in the central region because of the inner boundary of the simulation.
{Alt text: Four color and contour images in the left column, and four line graphs in
the right column. In the right column, x axis shows the positional offset from the protostellar
    position from -50 to 50 astronomical unit. The y axis shows the brightness temperature
    from 0 Kelvin to 250 Kelvin.}}
    \label{fig:Tbmap_m1}
\end{figure*}

The results of the radiative transfer calculation are shown in Figure \ref{fig:Tbmap_m1}.
The left panels show the brightness temperature images at 1.3 mm, and the right panels show the brightness temperature profile along the major axis shown by the dashed lines in the left panels.
The results of the face-on view without beam convolution, the face-on view with convolution, the inclined view without convolution, and the inclined view with convolution, are presented in each row from
top to bottom.
Spiral structures are seen in the panels without inclination. 
On the contrary, after the inclination the spiral features are
only margninaly discernible. In particular,
the case of the inclined disk with convolution (the fourth row), which should be the representative of the actual observations,
does not clearly exhibit a sign of the spiral structure.
These results imply that the effect of the beam convolution and the inclination makes the spiral structure undetectable in the eDisk observations. 
The apparent absence of spiral features in the eDisk 1.3-mm image in IRAS 16544 thus does not necessarily indicate the real absence of spiral structures and gravitational instability in the disk.

The intensity profiles along the major axis show
hints of the internal structures of the disk.
The intensity profile of the face-on,
no convolution case presents a number of local emission maxima
originated from the spiral features. Whereas these local emission
maxima are smoothed out by the effects of the beam convolution
and inclination, the intensity profiles in the other three cases still show
bumps. In the case of the inclined,
convolved case, $i.e.,$ observational case, they are seen
as an almost symmetric shoulder structure on both side along the major axis.

\begin{figure}
    \includegraphics[width=8.8cm]{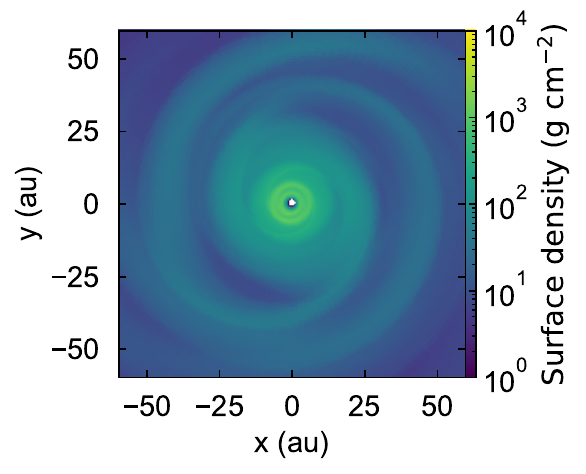}
    \includegraphics[width=8.8cm]{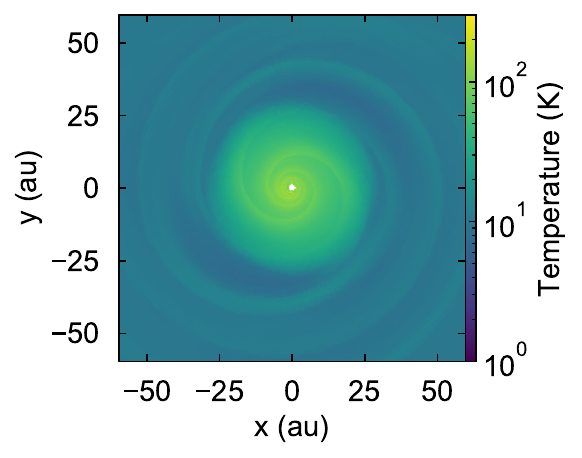}    
    \caption{Same as Figure \ref{fig:sim_results} but for Model 2.
    {Alt text: Two color images.}}
    \label{fig:sim_results2}
\end{figure}

\begin{figure*}
\begin{tabular}{cc}
\begin{minipage}[t]{0.36\hsize}
    \centering
    \includegraphics[keepaspectratio, scale=0.6]{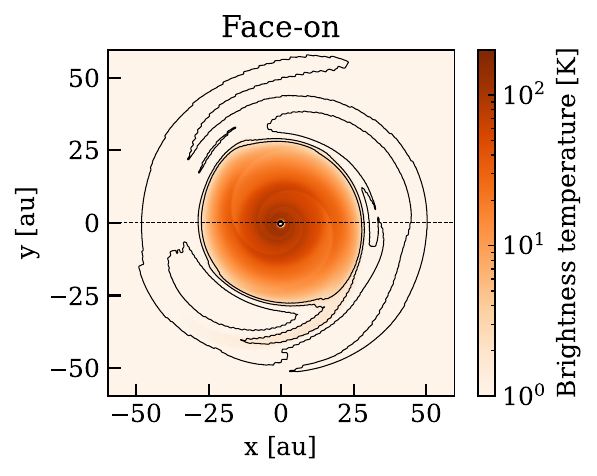}   
\end{minipage} &
\begin{minipage}[t]{0.54\hsize}
    \centering
    \includegraphics[keepaspectratio, scale=0.6]{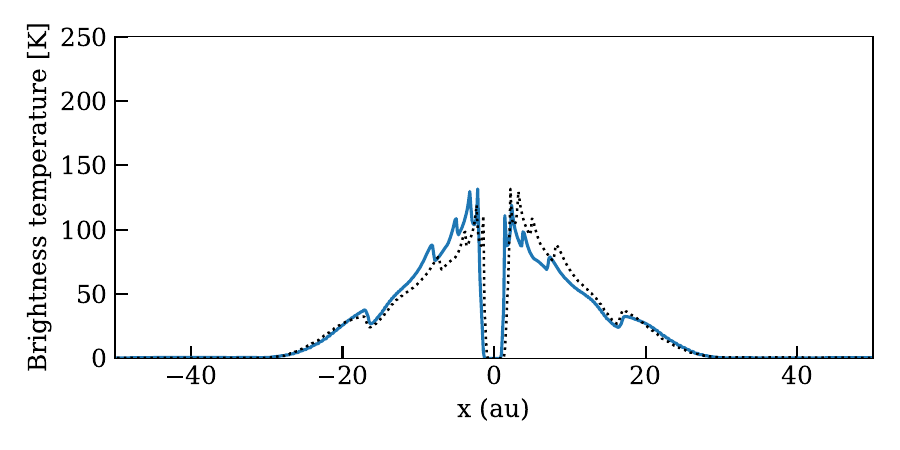}   
\end{minipage} \\
\hline

\begin{minipage}[t]{0.36\hsize}
    \centering
    \includegraphics[keepaspectratio, scale=0.6]{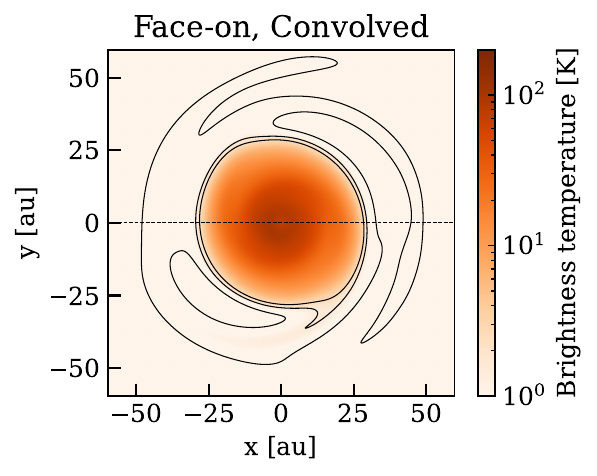}    
\end{minipage} &
\begin{minipage}[t]{0.54\hsize}
    \centering
    \includegraphics[keepaspectratio, scale=0.6]{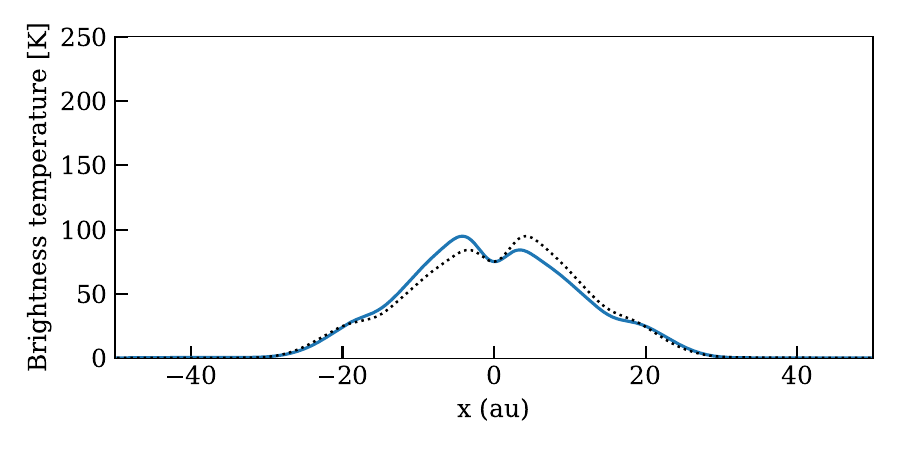}   
\end{minipage} \\
\hline

\begin{minipage}[t]{0.36\hsize}
    \centering
    \includegraphics[keepaspectratio, scale=0.6]{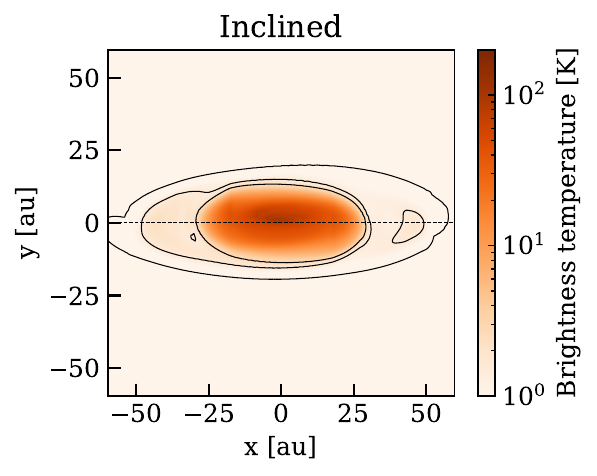}    
\end{minipage} &
\begin{minipage}[t]{0.54\hsize}
    \centering
    \includegraphics[keepaspectratio, scale=0.6]{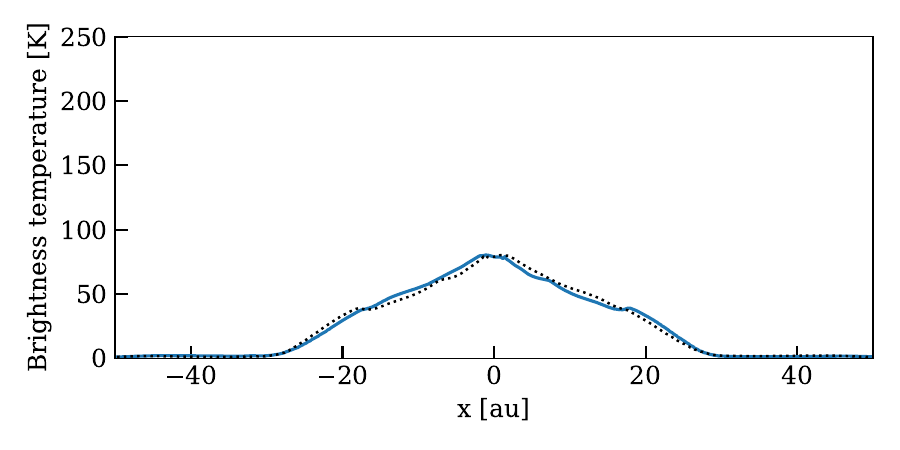}   
\end{minipage} \\
\hline

\begin{minipage}[t]{0.36\hsize}
    \centering
    \includegraphics[keepaspectratio, scale=0.6]{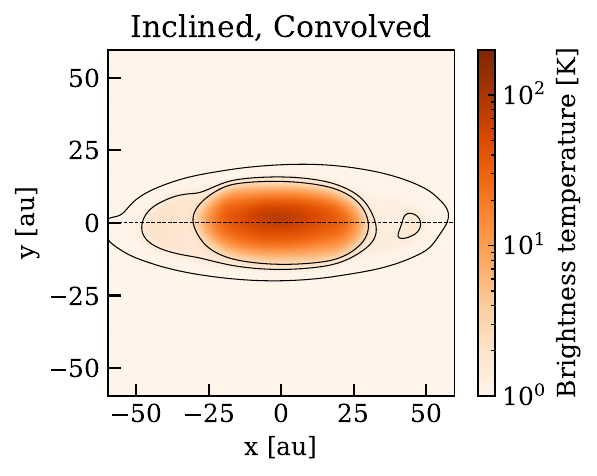}    
\end{minipage} &
\begin{minipage}[t]{0.54\hsize}
    \centering
    \includegraphics[keepaspectratio, scale=0.6]{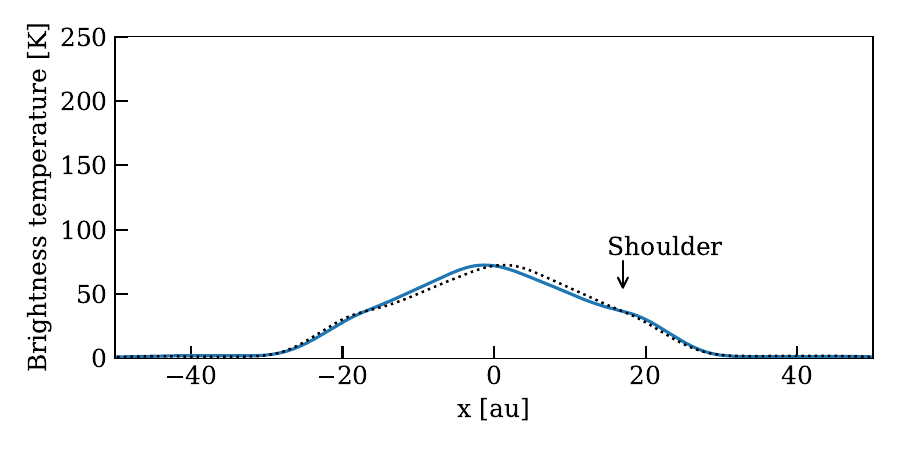}
\end{minipage} 
\end{tabular}
    \caption{{Same as Figure \ref{fig:Tbmap_m1} but for Model 2.
    Dotted lines in the left panels denote the intensity profile flipped horizontally,
    to show the asymmetry of the intensity profiles more clearly.
    A possible shoulder structure to the right hand side of the intensity profile is seen
    in the inclined and convolved case.}
    {Alt text: Four color and contour images in the left column, and four line graphs in
the right column. In the right column, x axis shows the positional offset from the protostellar
    position from -50 to 50 astronomical unit. The y axis shows the brightness temperature
    from 0 Kelvin to 250 Kelvin.}}
    \label{fig:Tbmap2}
\end{figure*}

\subsection{Models 2 and 3: High Surface Density Case}

\ \ \ \ The surface density and temperature distributions obtained from the numerical simulation of Model 2, with the initial
disk gas mass of 0.17 $M_{\rm \odot}$ and the
surface density 1.75 times higher than that of Model 1,
are shown in Figure \ref{fig:sim_results2}.
The model images and intensity profiles along the major axis after the radiative transfer calculations are shown in Figure \ref{fig:Tbmap2}.
Similar to Model 1, spiral features are barely seen
in the inclined and real observational cases.
On the other hand, the flipped intensity profiles along the major axis
(black dotted lines in Figure \ref{fig:Tbmap2} right)
reveal an asymmetry. In particular, the real observational case
shows an asymmetric shoulder feature
(Figure \ref{fig:Tbmap2} bottom-right).

\begin{figure}
    \includegraphics[width=8.8cm]{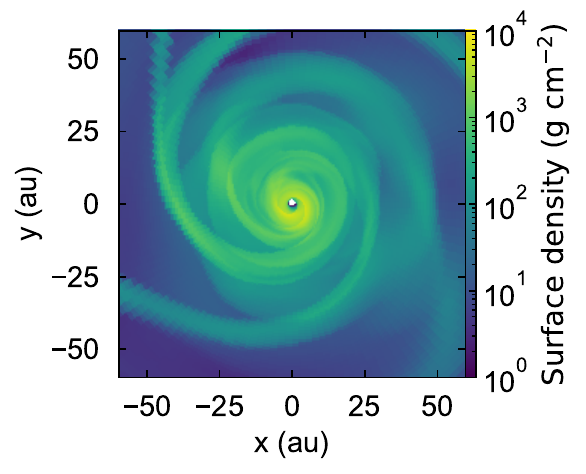}
    \includegraphics[width=8.8cm]{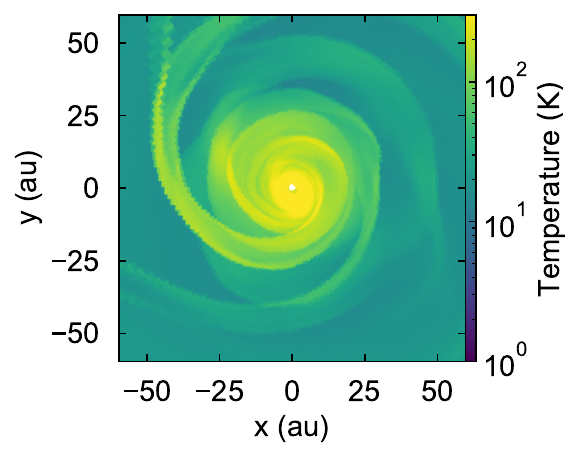}    
    \caption{Same as Figure \ref{fig:sim_results} but for Model 3
    at $t$ = 3.5 $\times$ 10$^3$ yr after the start of the simulation.
    {Alt text: Two color images.}}
    \label{fig:sim_results3}
\end{figure}

\begin{figure}
    \includegraphics[width=8.8cm]{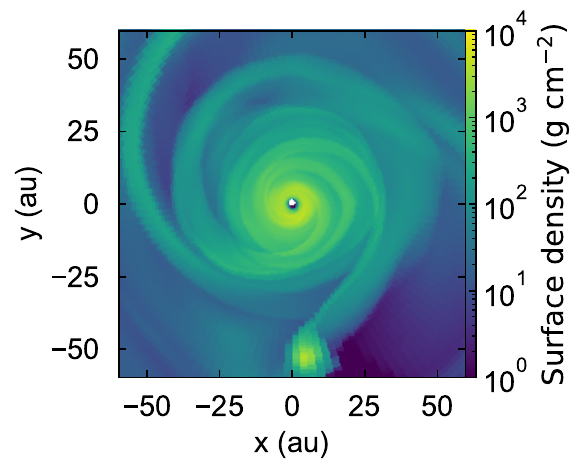}
    \includegraphics[width=8.8cm]{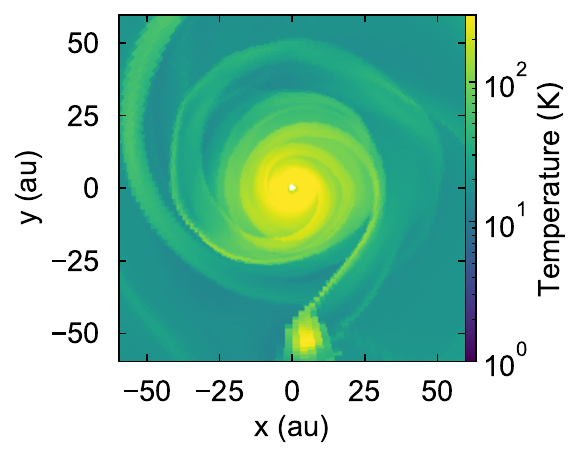}    
    \caption{Same as Figure \ref{fig:sim_results} but for Model 3
    at $t$ = 4.1 $\times$ 10$^3$ yr after the start of the simulation.
    {Alt text: Two color images.}}
    \label{fig:sim_results3_frag}
\end{figure}

\begin{figure*}
\begin{tabular}{cc}
\begin{minipage}[t]{0.36\hsize}
    \centering
    \includegraphics[keepaspectratio, scale=0.6]{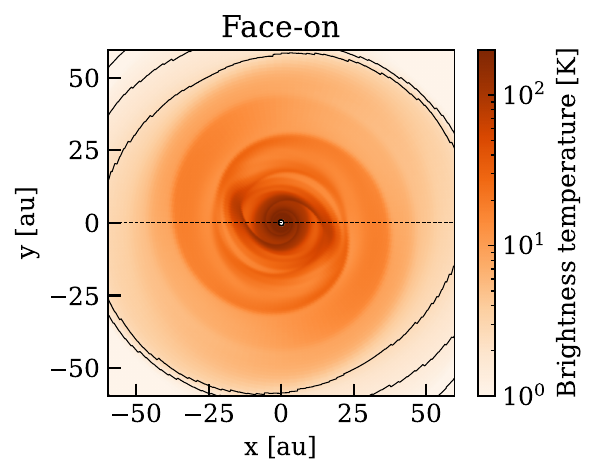}   
\end{minipage} &
\begin{minipage}[t]{0.54\hsize}
    \centering
    \includegraphics[keepaspectratio, scale=0.6]{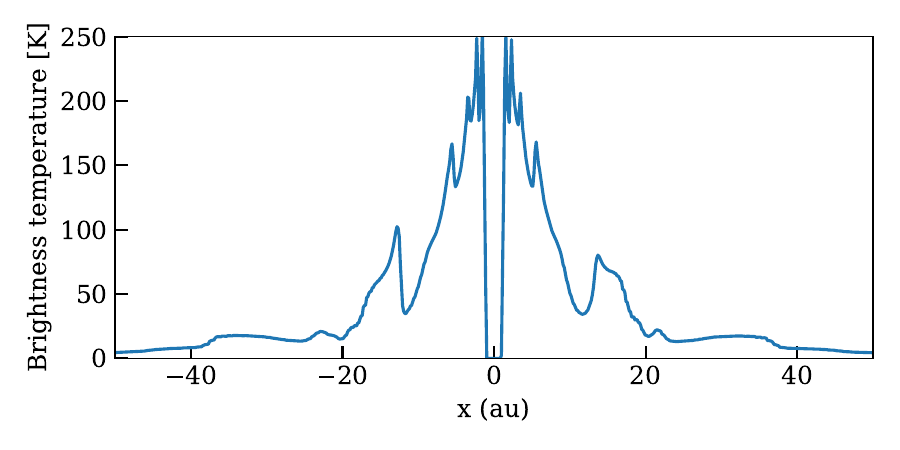}   
\end{minipage} \\
\hline

\begin{minipage}[t]{0.36\hsize}
    \centering
    \includegraphics[keepaspectratio, scale=0.6]{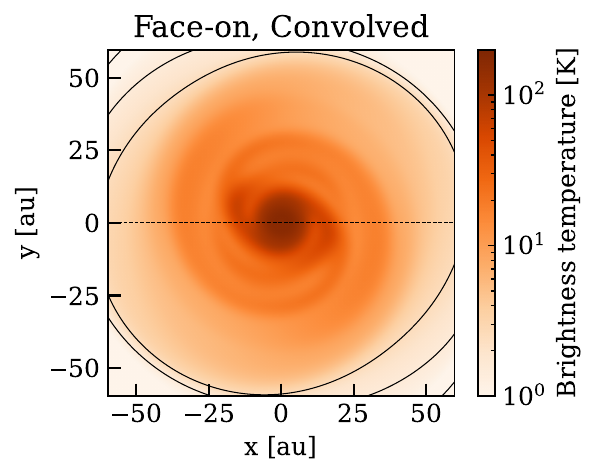}    
\end{minipage} &
\begin{minipage}[t]{0.54\hsize}
    \centering
    \includegraphics[keepaspectratio, scale=0.6]{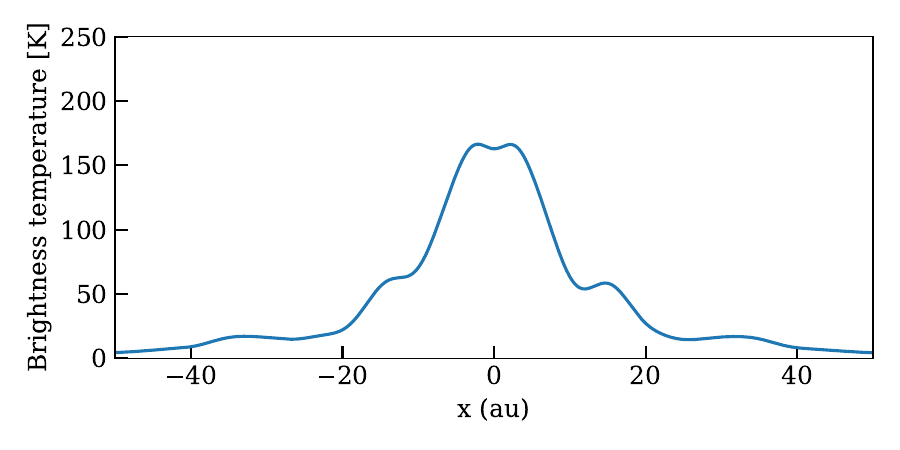}   
\end{minipage} \\
\hline

\begin{minipage}[t]{0.36\hsize}
    \centering
    \includegraphics[keepaspectratio, scale=0.6]{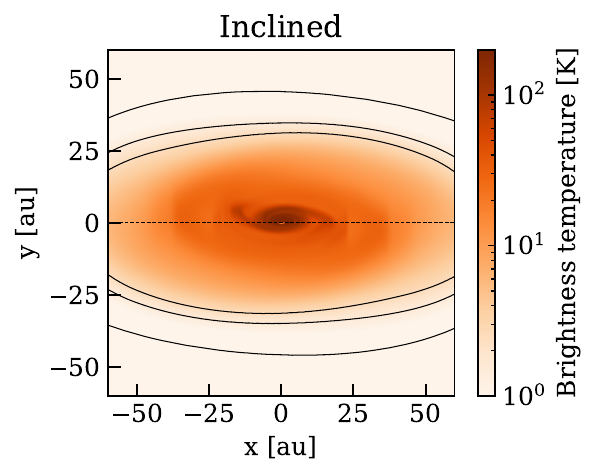}    
\end{minipage} &
\begin{minipage}[t]{0.54\hsize}
    \centering
    \includegraphics[keepaspectratio, scale=0.6]{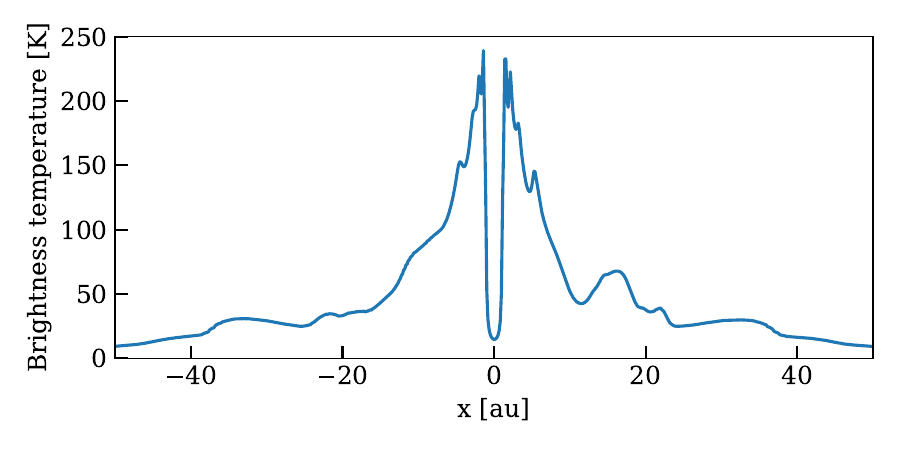}   
\end{minipage} \\
\hline

\begin{minipage}[t]{0.36\hsize}
    \centering
    \includegraphics[keepaspectratio, scale=0.6]{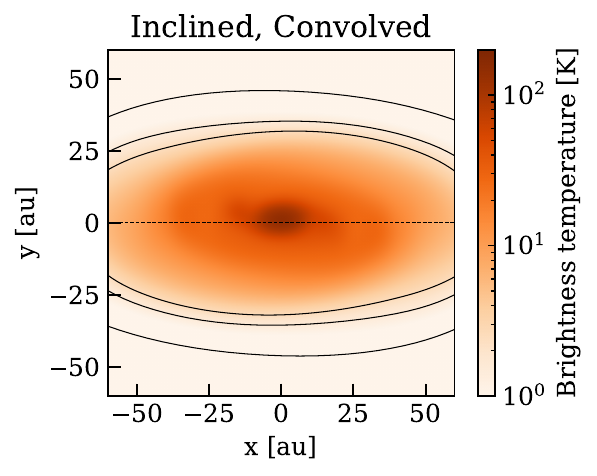}    
\end{minipage} &
\begin{minipage}[t]{0.54\hsize}
    \centering
    \includegraphics[keepaspectratio, scale=0.6]{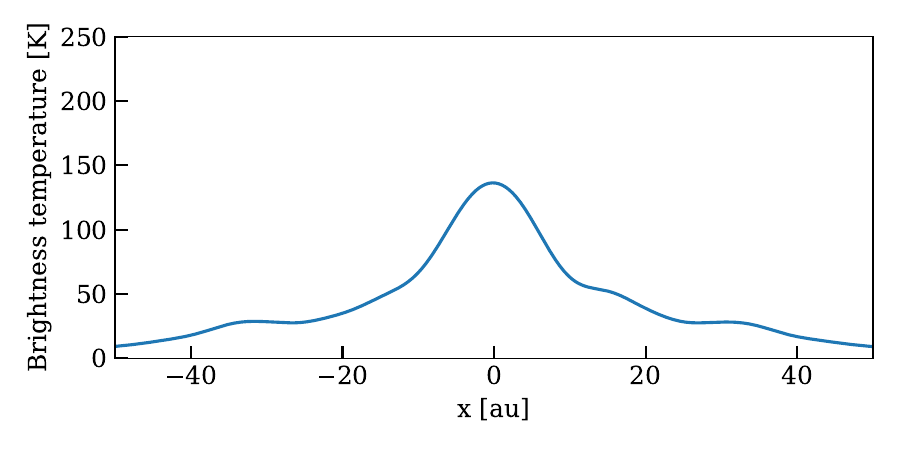}
\end{minipage} 
\end{tabular}
    \caption{Same as Figure \ref{fig:Tbmap_m1} but for Model 3
    at $t =$3.5 $\times$ 10$^3$ yr.
    {Alt text: Four color and contour images in the left column, and four line graphs in
the right column. In the right column, x axis shows the positional offset from the protostellar
    position from -50 to 50 astronomical unit. The y axis shows the brightness temperature
    from 0 Kelvin to 250 Kelvin.}}
    \label{fig:Tbmap3}
\end{figure*}

The surface density and temperature distributions of Model 3
at $t =$3.5$\times$10$^3$ yr and $t =$4.1$\times$10$^3$ yr
are shown in Figures \ref{fig:sim_results3} and \ref{fig:sim_results3_frag},
respectively. This is the case of the most massive disk
with the initial disk gas mass of 0.21 $M_{\rm \odot}$ and the
surface density more than twice higher than that of Model 1.
The model exhibits a three-arm, $m = 3$ mode feature
in the outer part of the disk both in the density and temperature maps
at $t =$3.5$\times$10$^3$ yr.
After $t =$3.9$\times$10$^3$ yr the disk fragments into
two components. Since the observed disk does not
show such a fragment feature, the earlier-time result is adopted
for the radiative transfer calculation.
The model images and intensity profiles along the major axis
after the radiative transfer calculations of Model 3
at $t =$3.5$\times$10$^3$ yr
are shown in Figure \ref{fig:Tbmap3}.
The $m = 3$ pattern creates a clear asymmetric
intensity profile along the disk major axis, particularly evident
in the non-convolved cases (Figure \ref{fig:Tbmap3}).
Even in the inclined and convolved case, the intensity profile along
the major axis shows a clear asymmetry, and there is a shoulder feature
at $r \sim$15 au to the righthand side
(bottom-right panel of Figure \ref{fig:Tbmap3}).

\begin{figure*}
    \includegraphics[width=16cm]{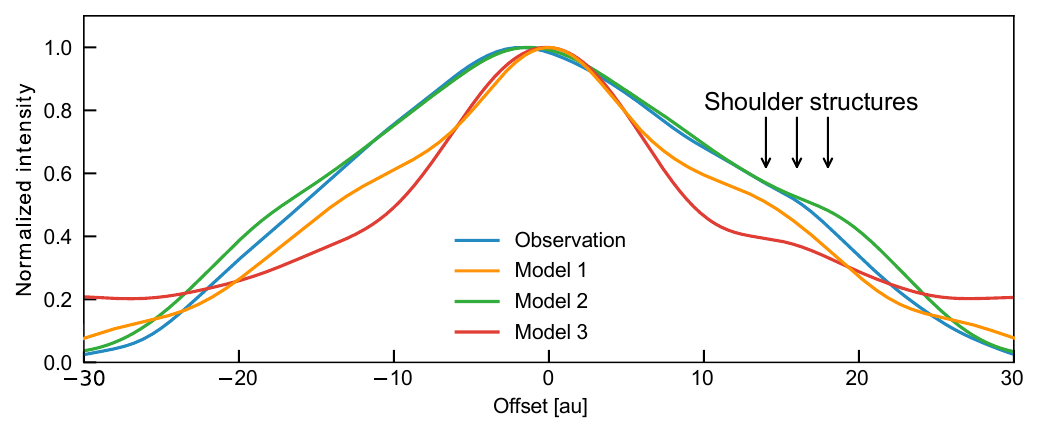}
    \caption{Normalized intensity profiles along the major axis in the observations (blue), Model 1 (orange), Model 2 (green), and
    {\bf Model 3 (red)}.
    Horizontal axis shows the offset from the center,
    where the positive offset corresponds to the side with the shoulder structure.
    {Alt text: Line graph. x axis shows the positional offset from the protostellar
    position from -30 to 30 astronomical unit. The y axis shows the normalized intensity from 0 to 1.1.}
    }
    \label{fig:Tbnorm}
\end{figure*}

Figure \ref{fig:Tbnorm} compares the observed and Models 1, 2, and 3
intensity profiles along the major axis, where the intensity
scales are normalized with the peak intensities for a direct
comparison of the profile shapes. All the models
reveal a shoulder feature at similar radii to that of
the observed shoulder feature. Model 2 further reproduces
the observed overall asymmetric shape of the profile with the peak
position offset from the origin.
The disk of Model 3 appears most extended toward the outer part,
while the extent of Model 1 is the smallest. This follows
the degree of the gravitational instability, where under a weaker
gravitational instability gas tends to accumulate toward the center.

These results imply that spiral arms created by the gravitational
instability (GI) could be the source of the shoulder and asymmetric
features observed in the protostellar disks.

\section{Discussion}\label{sec:discussion}
\begin{figure*}
    \includegraphics[width=8cm]{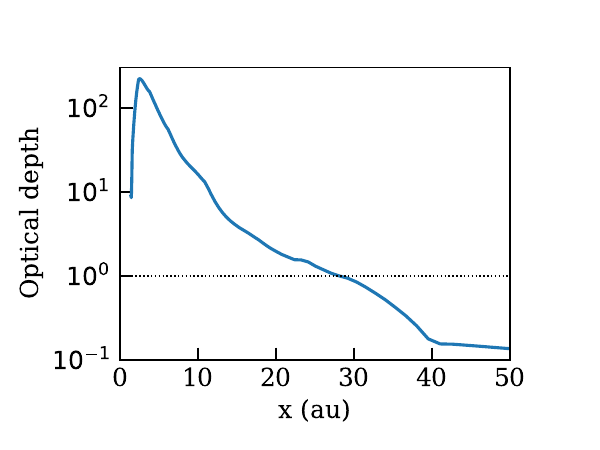}   
    \includegraphics[width=8cm]{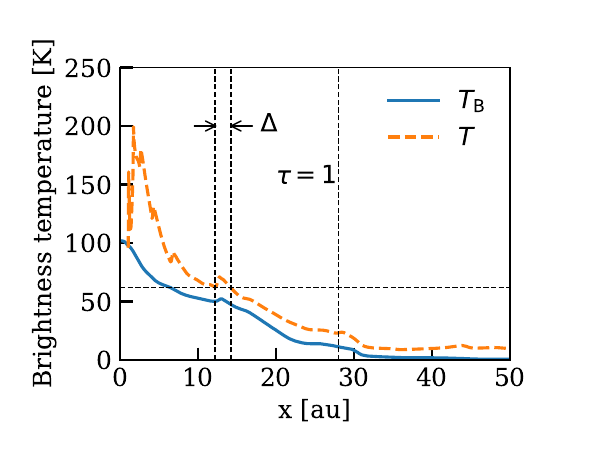}
    \caption{Radial profile of the optical depth at 1.3 mm (left panel) and
    that of the 1.3-mm brightness temperature and the dust temperature of the disk model (right panel) along the major axis of Model 1. The rapid decrease of the profiles
    around the center is the artifact of the inner boundary of the numerical simulation. In the right panel, the brightness temperature and the dust temperature are shown by solid blue and dashed orange lines, respectively. The width of the spiral $\Delta$ is defined by the spiral region where the temperature is higher than the closest local minima (see the horizontal dashed line).
    The vertical dashed line at $x=28$ au shows the radius where $\tau =1$.
    {Alt text: Two line graphs. x axis shows the radius from 0 to 50 astronomical unit.}}
    \label{fig:sim_x}
\end{figure*}

\subsection{Required Resolution and Sensitivity to Reveal the Spirals}
\label{beamsize}

\ \ \ \ As demonstrated above, the effects of the beam convolution and
disk inclination hide the spiral structures created by GI.
In the case of the disk in IRAS 16544 with an inclination angle of
$\sim$73$^\circ$, the projected spatial scale is shrunk by
$\sim$30 \% along the minor axis.
It is thus required to investigate the intensity profile along the major axis to study internal structures in the disk.
Figure \ref{fig:sim_x} (left) shows the radial profile of the dust optical depth at 1.3 mm along the major axis in Model 1
(inclined view without convolution; the third row of Figure \ref{fig:Tbmap_m1}). Over almost the entire radius of the observed dust
disk ($r_{\rm dust} = 30$ au), the 1.3-mm dust-continuum emission
is optically thick. This implies that the spiral structures
seen in the model images reflect the temperature enhancements
due to the compressional heating in the spirals.
Figure \ref{fig:sim_x} (right) shows the relevant
model radial profiles of the brightness and dust temperatures.
Note that the brightness temperature profile is affected
by the disk inclination while the dust temperature is not.
The brightness temperature (blue line) is lower than the dust temperature
(orange dashed line),
because the brightness temperature of the inclined disk with a certain thickness is affected by the outer, colder dust located on the front side of the line of sight.
In addition, the detailed structure in the dust temperature distribution within $r\sim$10 au is smoothed out
in the brightness temperature distribution because of the inclination
effect.
Only the substructures at $r \sim$13 au and $\sim$30 au originated
from the spirals
can be identified in the brightness temperature profile.



\begin{figure}
    \includegraphics[width=8cm]{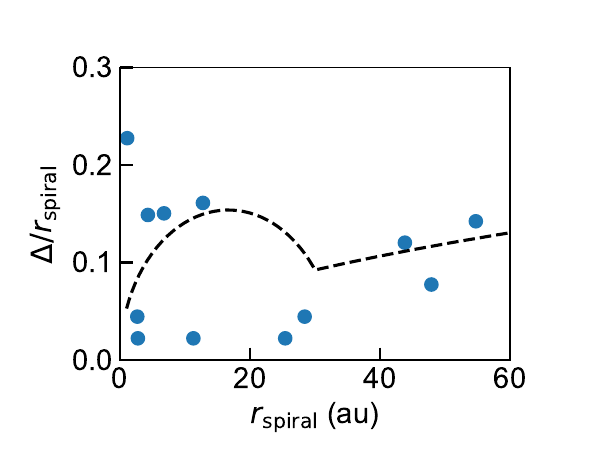}
    \caption{Radial distribution of the spiral width $\Delta$ normalized by the radius of the spiral $r_{\rm spiral}$ calculated from the dust temperature profile along the major axis (the right panel of Figure \ref{fig:sim_x}). The dashed line shows the pressure scale height normalized by the radius calculated from
    Equation \ref{eq:Hmodel}, where the temperature profiles given in
    Equation \ref{eq:Teq} is adopted.
    {Alt text: Line graph and plot of the data points.
    x axis shows the radius from 0 to 60 astronomical unit.
    y axis shows the normalized radius of the spiral location from 0 to 0.3.}}
    \label{fig:dr_spiral}
\end{figure}

\begin{figure}
    \includegraphics[width=8cm]{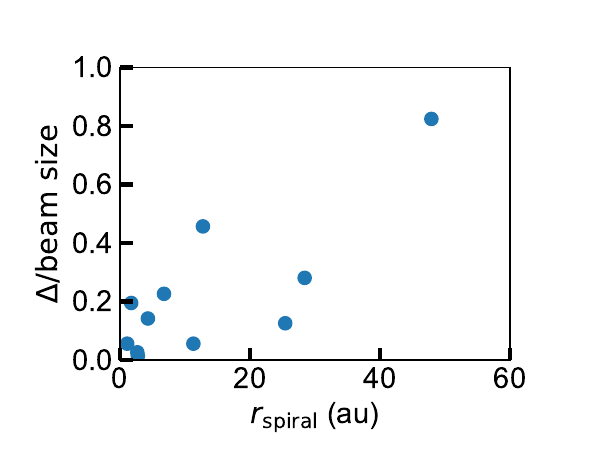}
    \caption{Radial distribution of the width of the spiral $\Delta$ normalized by the beam size of the IRAS 16544 observations.
    {Alt text: Plot of the data points.
    x axis shows the radius from 0 to 60 astronomical unit.
    y axis shows the normalized spiral width from 0 to 1.}}
    \label{fig:dr_perbeam}
\end{figure}

The observational beam size should be small enough to resolve these substructures.
Here the spiral radius $r_{\rm spiral}$ is defined as the
radius of the local maxima of the dust temperature.
The minimum dust temperature of the spiral is defined from the
intensity of the local minima closest to the local maxima,
and the spatial coverage above that local minima is defined as the
width of the spirals $\Delta$.
Figure \ref{fig:dr_spiral} shows the radial distribution of
$\Delta$ normalized over the spiral radius $r_{\rm spiral}$.
$\Delta$ is $\lesssim 0.1 r$ in half of the data points. In addition,
$\Delta$ is similar to the pressure scale height (dashed line in Figure \ref{fig:dr_spiral}),
which is a typical length scale of the self-gravitational instability.
Figure \ref{fig:dr_perbeam} shows the radial distribution of
$\Delta$ normalized by the beam size of the eDisk observations of
$\sim$4.5 au.
In all the spirals over the dust disk radius of 30 au, $\Delta$ is smaller than the beam size. To unveil the spiral structure
$\Delta$ should be sampled at least three independent beams,
implying that the a factor $\gtrsim$10 finer beam size of $\lesssim$0.5 au
is required.

%

\begin{figure*}
\begin{tabular}{cc}
\begin{minipage}[t]{0.36\hsize}
    \centering
    \includegraphics[keepaspectratio, scale=0.6]{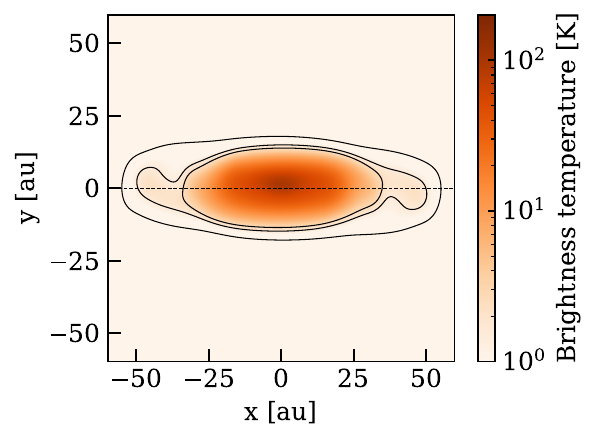}   
\end{minipage} &
\begin{minipage}[t]{0.54\hsize}
    \centering
    \includegraphics[keepaspectratio, scale=0.6]{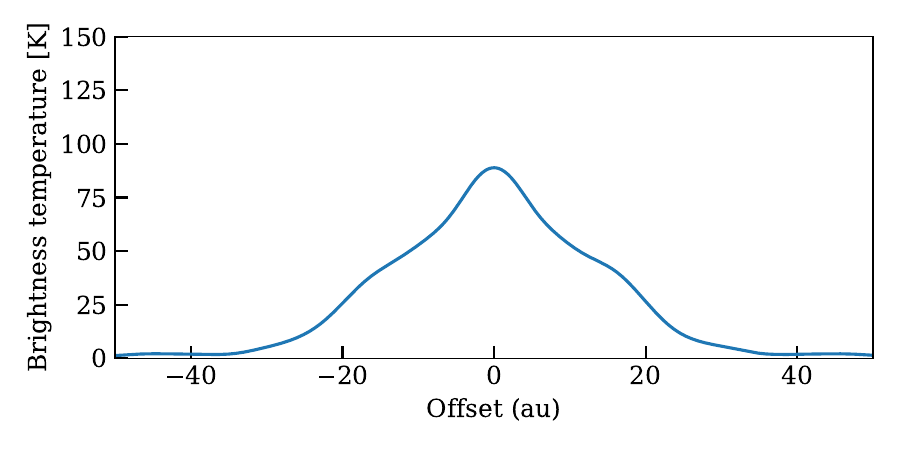}   
\end{minipage} \\

\begin{minipage}[t]{0.36\hsize}
    \centering
    \includegraphics[keepaspectratio, scale=0.6]{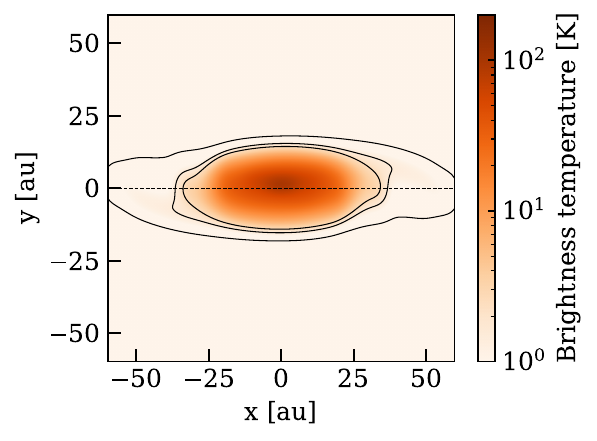}   
\end{minipage} &
\begin{minipage}[t]{0.54\hsize}
    \centering
    \includegraphics[keepaspectratio, scale=0.6]{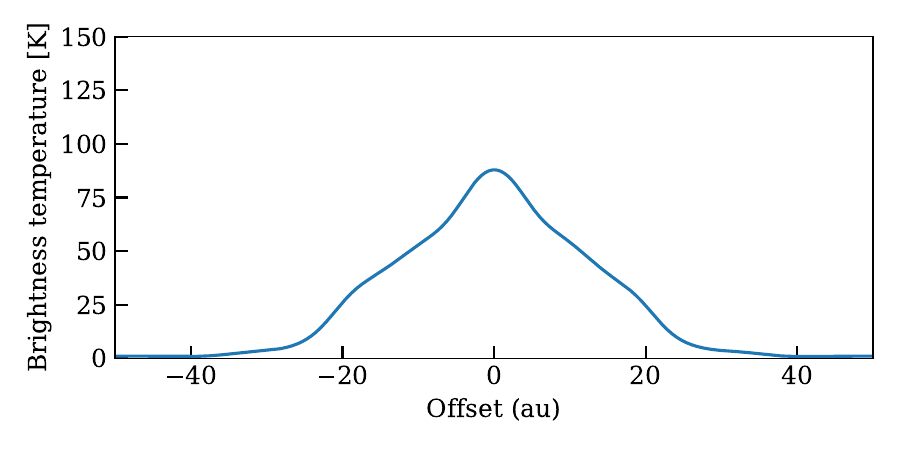}   
\end{minipage} \\

\begin{minipage}[t]{0.36\hsize}
    \centering
    \includegraphics[keepaspectratio, scale=0.6]{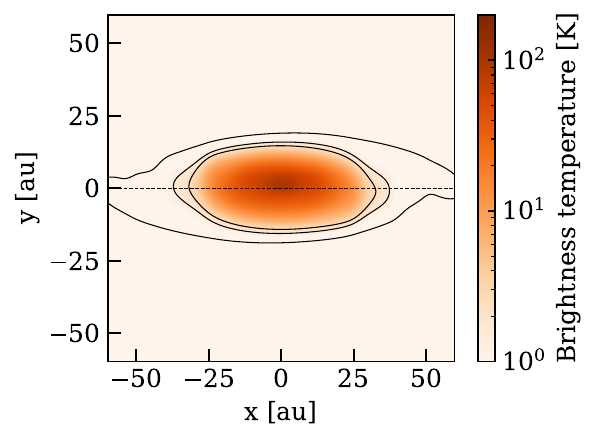}   
\end{minipage} &
\begin{minipage}[t]{0.54\hsize}
    \centering
    \includegraphics[keepaspectratio, scale=0.6]{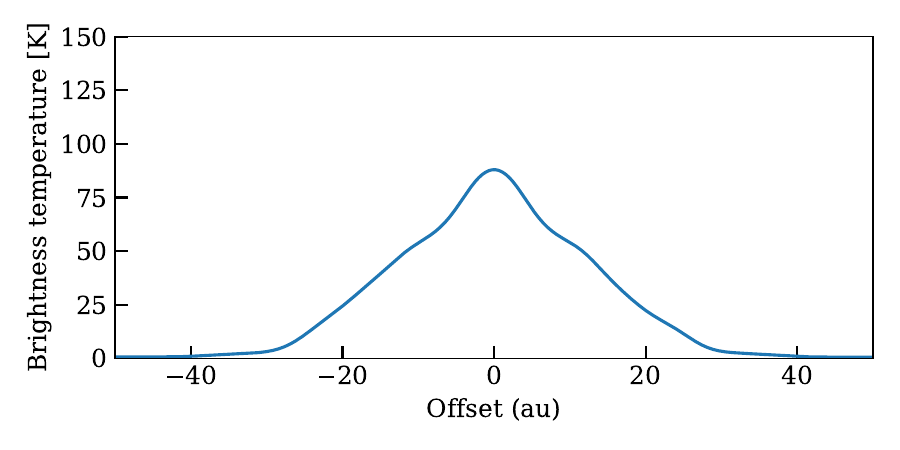}   
\end{minipage} \\
\end{tabular}
\caption{
Inclined and beam-convolved
model images (left panels) and intensity profiles along the major axis (right) with a different azimuthal angle of the disk major axes.
These correspond to the bottom panels of Figure \ref{fig:Tbmap_m1} but
the azimuth angle is rotated by 45$^\circ$, 90$^\circ$, and 135$^\circ$
proceeding from top to bottom. The contour levels in left panels are 1$\sigma$, 3$\sigma$, and 5$\sigma$ of the eDisk observations (1$\sigma$= 0.5 K).
{Alt text: Three color and contour images in the left column, and three line graphs in
the right column. In the right column, x axis shows the positional offset from the protostellar
    position from -50 to 50 astronomical unit. The y axis shows the brightness temperature
    from 0 Kelvin to 150 Kelvin.}}
\label{fig:pa}
\end{figure*}

To investigate the effect of the direction of the azimuthal cuts,
Figure \ref{fig:pa} shows the inclined, beam-convolved images and intensity profiles of Model 1
for different azimuthal directions of the disk major axes.
Depending on the azimuthal directions, a spiral feature in the outer
part of the disk appears in the 1 and $3\sigma$ contours, but not
in the $5\sigma$ contour.
Thus, $\gtrsim$2 times longer observing time
is also required to detect the outer spirals.
Note that symmetric shoulder-like structures are evident in the right panels
of Figure \ref{fig:pa} and their radii vary depending on
the observing direction.
This result implies that the position of the shoulder structure formed by the spirals depends on the observation timing of the orbital motion.

\subsection{Flared Dust Distribution in Young Protostellar Disks}
\label{flared}

\ \ \ \ {Our 2-dimensional simulations and radiative transfer
calculations assume that
dust grains of all sizes are flared following eqs. 9 and 10.
Our modeling thus does not consider the difference of the scale height
depending on the dust sizes. The use of Rosseland mean opacity
to calculate the dust temperature in the 2-dimensional
\textsc{fargo-adsg} simulations implicitly assumes the
vertically uniform dust distribution regardless of the dust size.
The 1.3-mm dust-continuum emission
is sensitive to sub-millimeter-size dust grains, which are considered
to be more settled onto the disk midplane than smaller grains,
as in the case of protoplanetary disks with ring and gap features
\cite[e.g.,][]{2016ApJ...816...25P,2022ApJ...930...11V}.
If such dust settling is present in IRAS 16544, our simulations
and radiative transfer calculations could over-estimate
the observability of the spiral features.

Our eDisk observations of young protostellar disks, however,
show that large dust grains traced by the 1.3-mm continuum emission
are likely flared. Several of eDisk targets, including IRAS 16544, exhibit
asymmetric intensity profiles along the disk minor axes.
Our radiative transfer modeling demonstrates that such asymmetric
intensity profiles can be interpreted as the flaring of large dust
and the optically-thick dust-continuum emission \citep{2024ApJ...964...24T}.
Furthermore, in gravitationally unstable disks, turbulent viscosity
{\bf
\begin{equation}
\nu_{\rm vis} = \alpha c_s H
\end{equation}
}
is considered to be high because of a large effective viscous parameter
$\alpha$ ($\geq$0.01) inferred from hydrodynamic simulations
\citep[e.g.,][]{2001ApJ...553..174G,2005MNRAS.364L..56R}.
Such a high turbulent $\alpha$ is sufficient to maintain
a high scale height of sub-mm-sized dust grains.
Therefore, in the case of younger, protostellar disks,
flared distribution of sub-mm-sized dust grains is expected
and the present 2-dimensional \textsc{fargo-adsg} simulations
along with the radiative transfer calculations should be valid
to discuss the observability of the internal spiral features.
}

\subsection{Implications for Formation and Evolution of Disk Internal Structures}
\label{comp}

\ \ \ \ Our numerical simulations along with the radiative transfer calculations demonstrate that a shoulder feature
can be reproduced by spiral arms in a marginally gravitationally unstable
($Q \sim$2) disk model (Model 1). In a more massive, gravitational
unstable disk (Model 2; $Q \sim$1.3),
an asymmetric feature along the disk major axis can also be created.
In the disk more massive than the central protostar (Model 3; $Q \sim$1),
a three-arm, $m = 3$ mode can also be produced, which exhibits a clear
asymmetric shoulder feature in the intensity profile along the major axis.

Similar asymmetric bump features are seen toward Ced110 IRS4A, and intriguing western, central, and eastern peaks are seen in the disk toward Ced110 IRS4B \cite[]{2023ApJ...954...67S}. In L1527 IRS, the southern part of the disk is clearly stronger than the northern part \cite[]{2023ApJ...951...10V}. Class 0 sources of IRAS 16253-2429 \cite[]{2023ApJ...954..101A}, R CrA IRS5N \cite[]{2023ApJ...954...69S}, and a Class I source R CrA IRS7B-a \cite[]{2024ApJ...964...24T}, show the skewed intensity peaks along the disk major axes.
Except for R CrA IRS7B-a, sources with such disk substructures are all Class 0 sources associated with Keplerian rotations. In the case of an even younger source of B335, a clear disk like signature in the 1.3-mm dust-continuum emission and the Keplerian rotation are not identified \cite[]{2015ApJ...812..129Y, 2023ApJ...951....8O}. On the other hand, most evolved Class I sources in our eDisk sample, L1489 IRS and Opt IRS63, exhibit ring-gap features, while the image contrasts between the rings and gaps are not as evident as those in the Class II sources. These results suggest that the shoulder, bump, and/or asymmetric peaks may be common features in the Class 0 disks.

Our results indicate that these observations do not conflict with
previous theoretical studies of disk formation and evolution processes, which show that young disks are massive enough to be gravitationally unstable and to form spiral arms \cite[][]{2010ApJ...718L..58I,2018MNRAS.473.4868Z, 2020ApJ...896..158T}.
This is also important to understand formation mechanism of
ring-gap structures observed in more evolved disks than those observed by eDisk.
One of the commonly accepted mechanisms of ring-gap formation is the
carving by the hidden planet \cite[][]{2023ASPC..534..685P}.
The observed gap radii, however, reaches $\sim$100 au \cite[][]{2023ASPC..534..423B}, where it is difficult to form the planet
by the core accretion within the timescale of the disk age
($\lesssim 10^6$ yr).
Thus, gravitational fragmentation of the disk \cite[]{1997Sci...276.1836B} is one of the possible mechanisms to form the gap-forming planet because the timescale of the gravitational collapse is comparable to the dynamical timescale of the disk \cite[][]{2020MNRAS.493L.108B}.
The disk self-gravity can also produce ring-gap structures without hidden planets.
Previous study indicates that ring-gap structures can be formed by the secular gravitational instability \cite[]{2014ApJ...794...55T,2016AJ....152..184T,2018PASJ...70....3T,2020ApJ...900..182T}.
While this instability can grow even if the disk is gravitationally stable, self-gravity can further facilitate this instability.
Thus, the prevalence of the ring-gap feature in evolved disks
favors grativational instability in protostellar disks.
Because the mass infall rate from the envelope onto the disk decreases with time and the disk gas accretes onto the central star, the disk surface density will decrease with time evolution.
Thus, more evolved disks such as those observed by DSHARP are expected to be gravitationally stable even if the disks are gravitationally unstable in their protostellar disk stage.

At present, however, our study is a case-study of only one eDisk target,
IRAS 16544. In this particular case, the spiral arms due to gravitational instability, if present, cannot be imaged due to the effect of the beam size and the inclination. Further modeling and radiative transfer
calculations of a statistically significant sample of protostars are
required to discuss physical conditions and evolution of protostellar disks
into planet formation.

\section{Conclusions}\label{sec:summary}

\ \ \ \ We constructed numerical models of self-gravitating protostellar disks and performed radiative transfer calculations of the physical models. The model images exhibit spiral structures caused by the gravitational instabilities. We made detailed comparisons of those model images to the 1.3-mm image of the protostellar disk around the Class 0 protostar IRAS 16544 observed with eDisk. The effects of the disk inclination and the observational spatial resolution are incorporated, and
the origin of the shoulder feature as well as the apparent absence of the spiral features in the eDisk image is discussed. The findings of this work are summarized as follows:
\begin{itemize}
\item{
Spiral structures formed by the gravitational instability in the model disks are undetectable in the images after the effects of the inclination and beam convolution are incorporated.
This indicates that the non-detection of a clear spiral structure in the disk image
around IRAS 16544 does not reject the presence of gravitational instability and spiral features.
Since a young disk is expected to be gravitationally unstable by many, although not all, numerical studies
of disk formation and evolution, this result shows that the observations are not inconsistent with previous understandings of
protostellar disk evolution.
}

\item{Because the 1.3-mm dust-continuum emission in the model disk is optically thick over an almost entire
extent of the dust disk ($r\lesssim$30 au),
the intensity profile traces the temperature distribution. The width
of the spiral structure in the temperature distribution is measured,
and the required beam size to resolve the spiral structure is as small as
$\sim$0.5 au in the inner ($\lesssim$15 au) radii, which is $\sim$10 times finer than that of the eDisk observations.
In the outer radii, $\sim$twice longer observing time is
required to clearly image the spiral structure.
}

\item{The 1.3-mm eDisk image of the protostellar disk in IRAS 16544 shows an asymmetric shoulder structure in the intensity profile along the major axis.
We found that the asymmetric shoulder-like structure can
be reproduced when the disk is massive enough ($Q\lesssim1.3$).
In the model disk more massive than the central protostar,
a three-arm, $m = 3$ mode is also formed, which exhibits a clear asymmetric
shoulder structure in the intensity profile along the disk major axis.}
Shoulder-like or bump features are also observed in other eDisk samples, mostly Class 0 sources. Those observational results may suggest self-gravitating disks around those protostars in the early stage of disk evolution, before the stage of gap and ring features
caused by the protoplanets.
\end{itemize}

\section*{ACKNOWLEDGEMENTS}
This paper makes use of the following ALMA data: ADS/JAO.ALMA \#2019.1.00261.L and 2019.A.00034.S.
ALMA is a partnership of ESO (representing its member states), NSF (USA), and NINS (Japan), together with NRC (Canada), NSTC and ASIAA (Taiwan), and KASI (Republic of Korea), in cooperation with the Republic of Chile. The Joint ALMA Observatory is operated by ESO, AUI/NRAO, and NAOJ.
Numerical computations and analyses were in part carried out on Small Parallel Computers and analysis servers at Center for Computational Astrophysics, National Astronomical Observatory of Japan.
S.T. is supported by JSPS KAKENHI grant Nos. JP21H00048, JP21H04495, and
JP24K00674,
and by NAOJ ALMA Scientific Research grant No. 2022-20A. K.S. is supported by JSPS KAKENHI grant No. JP21H04495 and by NAOJ ALMA Scientific Research grant No. 2022-20A.

\section*{Appendix A: Radiative Transfer Modeling of the Dust Disk
of IRAS 16544-1604}

To investigate the effect of the flared disk structure
on the determination of the central protostellar position
and the major and minor axes through the 2-dimensional
Gaussian fitting, radiative transfer calculations
with RADMC3d have been conducted. The physical model
of the protostellar disk and envelope is the same
as that described by \citet{2024ApJ...964...24T}, $i.e.,$
a Keplerian disk with a vertically hydrostatic equilibrium
plus a rotating and infalling envelope taken from
\citet{1976ApJ...210..377U,1994ApJ...430L..49H,2004RMxAA..40..147M}.
Several of the model parameters are tailored to the case of
IRAS 16544 (Table \ref{tab:radparam}),
where the protostellar and disk mass, and the gas and dust disk radii
are identical to those of Model 1 in our \textsc{fargo-adsg} modeling.
Search for the other parameters was
made to approximately match the observed and model images
after the convolution with the observed beam.
Other model parameters not listed in Table \ref{tab:radparam}
are the same as those of the fiducial model
in \citet{2024ApJ...964...24T}.
As the purpose of the present radiative transfer calculations is
to investigate the effect of the flared disk structure on
the determination of the central protostellar position
and the major and minor axes, further fine tuning of the model parameters
to have an even better match of the model image with the observed image,
as well as interferometric observing simulations, is not performed.

\begin{figure*}
\includegraphics[width=20cm, trim=90 0 0 70, clip]{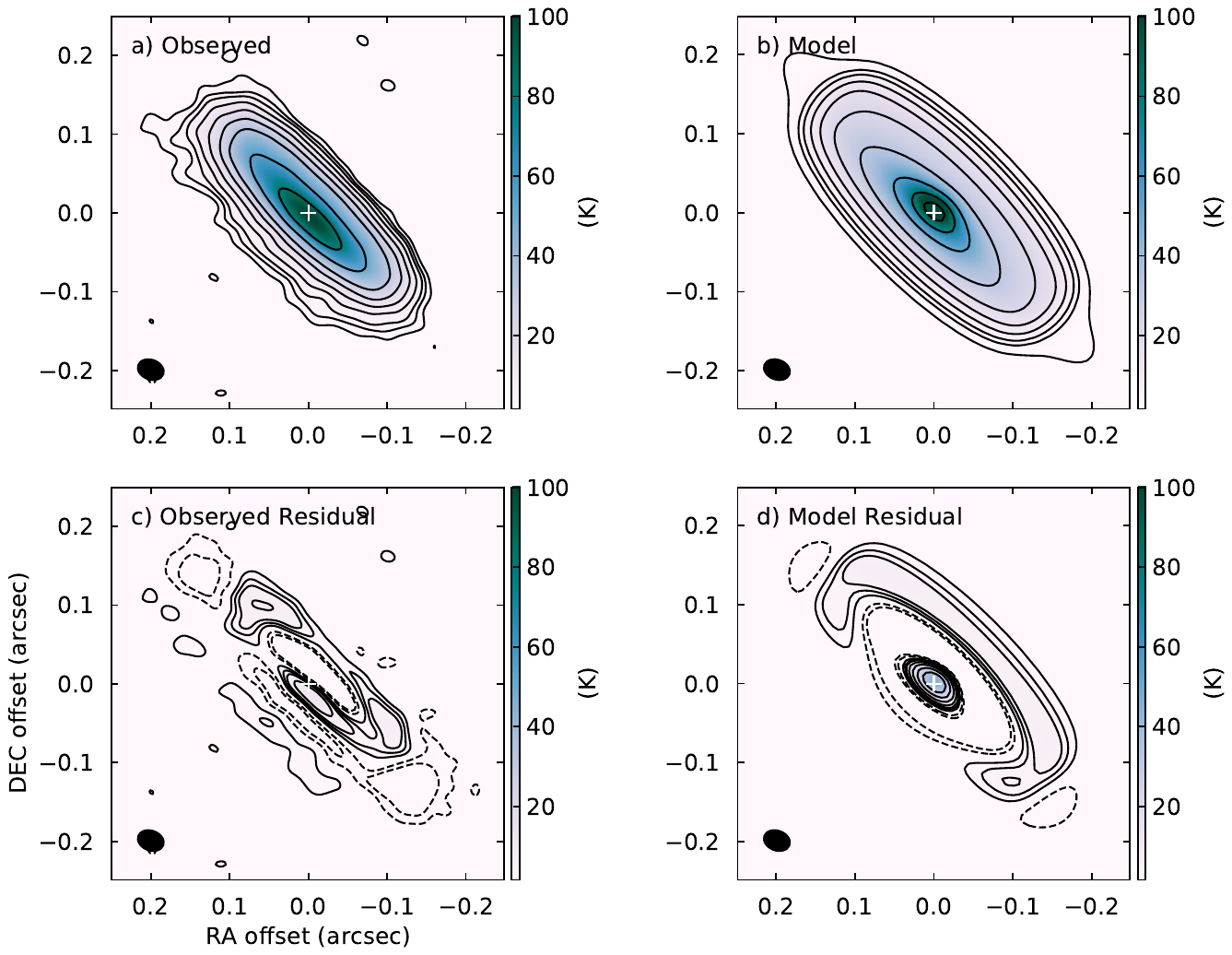} 
    \caption{Observed (a) and model images (b) of the 1.3-mm
    dust-continuum emission of IRAS 16544,
    and the residual observed (c)
    and model images (d) after the 2-dimensional Gaussian fitting.
    Contour levels are -6$\sigma$, -3$\sigma$, 3$\sigma$,
    6$\sigma$, 9$\sigma$, 15$\sigma$, 25$\sigma$, 40$\sigma$, 60$\sigma$,
    100$\sigma$, 150$\sigma$, 200$\sigma$ (1$\sigma$ = 0.522 K).
    Filled ellipses at the bottom-left corners denote the
    beam size (0$\farcs$0364 $\times$ 0$\farcs$0275; P.A. = 69.5$^\circ$).
    Alt text: Four color and contour images.
    }
    \label{fig:cb68gaussfits}
\end{figure*}

Figure \ref{fig:cb68gaussfits} shows the observed and model
images as well as the residual images after the fitting of
the 2-dimensional Gaussian fitting. 
The protostellar position of the model image is set to be
the centroid position as derived from the 2-dimensional
Gaussian fitting to the observed image
(16$^{\rm h}$57$^{\rm m}$19.6428$^s$, -16$^\circ$09\farcm24$\farcs$016;
white crosses in Figure \ref{fig:cb68gaussfits}).
The 2-dimensional Gaussian fitting to the model image
gives the Gaussian centroid position of
(16$^{\rm h}$57$^{\rm m}$19.6427$^s$, -16$^\circ$09\farcm24$\farcs$015),
which is indistinguishable from the central protostellar position.
The position angle as derived from the 2-dimensional Gaussian fitting
is 45.3$^\circ$$\pm$1.1$^\circ$, same as that of the correct answer
of 45$^\circ$. We thus consider that the effect of the flared disk structure
is not significant for the determination of the disk center and the major
and minor axes
through the 2-dimensional Gaussian fitting.

Both the observed and model residual images
show a positive, horseshoe feature to the northwest.
This asymmetric pattern in the residual images reflects
the asymmetric intensity profiles along the disk minor axis
due to the flared disk geometry \citep[][]{2024ApJ...964...24T}. 
Negative patterns to the northwest and southeast,
and residual peaks around the protostellar position,
are also seen, which show deviation of the intensity
profile from the Gaussian profile.

\begin{table*}[t]
\caption{Parameters of the radiative transfer modeling.}
\label{tab:radparam}
    \centering
\begin{tabular}{ll}
\hline
\hline
Protostellar mass                     & 0.14 $M_{\rm \odot}$\\
Protostellar luminosity               & 0.89 $L_{\rm \odot}$\\
Gas disk mass                         & 0.1 $M_{\rm \odot}$\\
Gas disk radius                       & 60 [au]\\
Dust disk radius                      & 30 [au] \\
Inclination angle                     & -73$^\circ$ \\
Position Angle                        & 45$^\circ$ \\
Disk flaring index $p$ ($H \sim r^p$) & 1.0 \\
Mass infalling rate                   & 2.0 $\times$ 10$^{-6}$ $M_{\rm \odot}~yr^{-1}$\\
Number of photons                     & 100000 \\
\hline
\end{tabular}
\end{table*}

{\it Software}: 
CASA \cite[]{2007ASPC..376..127M}, 
matplotlib \cite[]{2007CSE.....9...90H}, 
APLpy \cite[]{2012ascl.soft08017R,2019zndo...2567476R},
astropy \cite[]{2022ApJ...935..167A}, 
\textsc{fargo-adsg} \cite[]{2000A&AS..141..165M, 2008PhDT.......292B, 2008ApJ...678..483B}.

{\it Facility}: ALMA.

\bibliographystyle{aasjournal}
\bibliography{ref}{}

\end{document}